\documentclass[review,3p,times]{elsarticle}

\usepackage{amssymb}
\usepackage{amsmath}
\usepackage{booktabs}
\usepackage{subfigure}
\usepackage{multirow}
\usepackage{makecell}
\usepackage{nomencl}
\makenomenclature
\journal{arXiv}

\begin{document}

\begin{frontmatter}



\title{Virtual domain extension for imposing boundary conditions in flow simulation using pre-trained local neural operator}


\author[label1,label2]{Ximeng Ye\corref{cor1}} 
\ead{yeximeng@stu.xjtu.edu.cn}
\cortext[cor1]{Corresponding author}
          
\author[label3]{Hongyu Li} 
\ead{lihongyu_0@163.com}

\author[label1]{Zhen-Guo Yan} 
\ead{yanzhg@mail.ustc.edu.cn}

\affiliation[label1]{organization={State Key Laboratory of Aerodynamics},
             city={Mianyang},
            postcode={621000}, 
             state={Sichuan},
             country={PR China}}
\affiliation[label2]{organization={School of Energy and Power Engineering, Xi'an Jiaotong University},
             city={Xi'an},
             postcode={710049}, 
             state={Shaanxi},
             country={PR China}}
\affiliation[label3]{organization={China Aerodynamics Research and Development Center},
city={Mianyang},
postcode={621000}, 
 state={Sichuan},
 country={PR China}}

\begin{abstract}
This paper builds up a virtual domain extension (VDE) framework for imposing boundary conditions (BCs) in flow simulation using pre-trained local neural operator (LNO). 
It creates extended virtual domains to the input function to compensate for the corrosion nature of computational domains during LNO inference, thus turns the implementation of BC into the determination of field values on the extended domain. 
Several strategies to calculate the field values are proposed and validated in solving numerical examples, including padding operation, direct imposition, pressure symmetry, and optimization by backpropagation, and compared with boundary imposition in traditional solvers. 
It is found that the large time interval of LNO induces a relatively wide near-boundary domain to be processed, thus imposing BC on only a few nodes near the boundary following the immersed boundary conception in traditional solvers can hardly achieve high accuracy.
With appropriate values assigned on the extended virtual domains, VDE can accurately impose BCs and lead to reasonable flow field predictions. 
This work provides a guidance for imposing BCs reliably in LNO prediction, which could facilitate the reuse of pre-trained LNO in more applications.
\end{abstract}


\begin{keyword}
  local neural operator \sep boundary condition \sep reusability \sep virtual domain extension 



\end{keyword}

\end{frontmatter}



\section{Introduction}
\label{sec1}
Neural operators have recently made great success in solving partial differential equations (PDEs), featured in their ability of learning dynamic laws from high-resolution data and thus gaining higher performance compared to traditional numerical solvers \cite{Azizzadenesheli2023a}. 
The idea of neural operator is to train neural networks to approximate the mapping between functions, which is exactly the mapping needed for solving PDEs.
Representative methods include Fourier neural operator (FNO) \cite{Li2020b}, deep operator network (DeepONet) \cite{Lu2021b}, and their plentiful variants \cite{Li2021c,Wen2021,Rahman2022a,Haghighat2024,Howard2023,He2024}. 
As the next step, it seems very promising to exploit the advantage of neural operators to solve complex physical problems governed by PDEs.
However, to elevate these methods from simple benchmark problems to practical applications, the primary challenge is how to render one pre-trained neural operator reusable in multiple problems.

Reusability largely decides whether a neural network (NN) based method is competitive to traditional solvers. 
The data gathering and training process of NNs takes a considerable amount of time, which should be counted in when evaluating the efficiency.
The broader the reusable scope of the pre-trained NN, the higher the speed-up ratio, and thus the more deserving it is of meticulous architectural design and training.
Within the topic of solving transient PDEs, reusability of the pre-trained neural operator is mainly desired for different problem-specific conditions, including (1) the initial condition (IC), (2) the boundary condition (BC), and (3) the shape of the computational domain, of the same PDEs. 
The first factor, resuability on ICs, is inherently guranteed in neural operators.
Thanks to the definition of operator learning, a neural operator naturally accepts different input functions , which is exactly the reusability of different ICs for time-marching neural operators.
The reusability on the computational domain attracts a lot of attention since changing shapes is a common demand in engineering such as in the shape optimization process.
Several enhancement strategies have been reported, including simply parameterizing the shape (length, angle, and so on) \cite{Zhu2023} or encoding the geometric features by advanced NN techniques \cite{Cameron2023,Kashefi2022} as input, mapping the computational domain to a regular reference domain \cite{Gao2021,Li2023a,Yao2023,JMLR:v24:23-0064}, employing transfer learning \cite{Goswami2022}, and so on. 
Recently, we have proposed the local neural operator (LNO) \cite{Li2022c,Ye2023b} to improve the reusability at the neural operator's definition level. 
Two assumptions, the local-related and shift-invariant assumptions, are introduced to the neural operator definition. 
Then, the minimum input and output domains become small finite units that can move freely in space like a brush. 
This allows one pre-trained LNO to predict problems in various domains as any desired shape can be formed using these small units.

The remaining factor, the boundary condition, which is equivalently important as IC and the shape of computational domain, is less discussed. 
Imposition of boundary conditions largely influences the quality of predicted solutions since many common and complex phenomena in the solution of PDEs originate directly from boundary interactions (e.g., the boundary layer in fluid flow, the Rayleigh wave in the elastic free surface). 
However, the wide variety of boundary conditions, e.g., BCs of different formulations, with various values, and on various curves, makes them difficult to deal with in the prediction of pre-trained neural operators.
There are a few works touching on this topic. 
Refs. \cite{Wang2021i,Lotzsch2022,Tripura2022} take the boundary values as input variables, making one trained neural operator available for BCs of the same type but different values. 
Refs. \cite{Zhu2023,LuPestourie2021} modify the output of the neural operator to enforce the output as the desired value. 
Refs. \cite{Gao2021,Ren2021} use padding operation to deal with Dirichlet and Neumann BCs. 
These approaches do extend the reusable range, but only in a limited way. 
Ref. \cite{Yin2022} splits the domain into two parts to be predicted by the pre-trained neural operator and traditional finite element method (FEM) respectively, then the complex boundaries can be handled by FEM. 
This approach turns the imposition of boundary conditions in neural operators to that in traditional numerical theory, which is very effective, but reduces the superiority in efficiency. 
Our previous works \cite{Li2022c,Ye2023b,Ye2023} also made various attempts of imposing BCs including not only the aforementioned paddding operation and collaboration with FEM, but also immersed boundary method (IBM) to impose BCs on complex curves.
These attempts led to successful predictions, but they are more akin to an impromptu for each specific case rather than a general guidance.
Additionally, several questions remain: 
How accurate are these boundary treatments? Do they match the accuracy of the pre-trained LNO?
This work focuses on answering these questions and presents a systematic solution for imposing BCs.

\begin{figure}[htbp]
  \centering
  \includegraphics[width=0.85\textwidth]{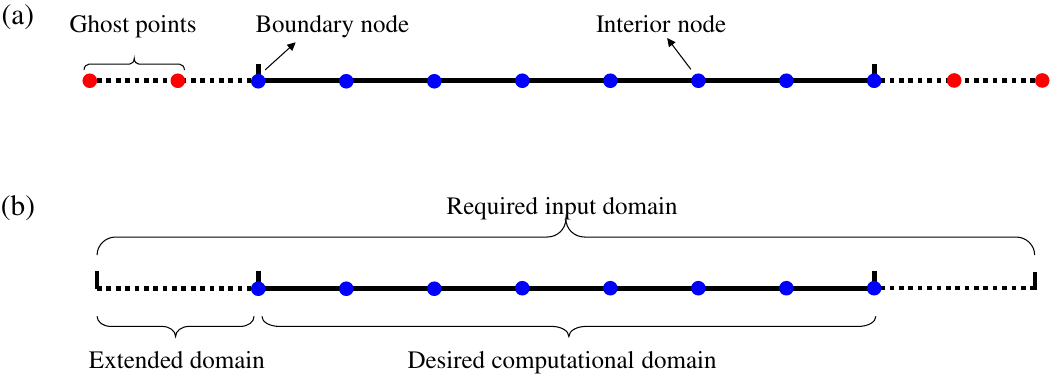}
  \caption{Comparison between the boundary treatment in FDM and LNO. (a) Ghost point in a 5-point central finite difference scheme. (b) Extended domain to predict the output on the desired computational domain in LNO.}\label{fig_ghost_point}
\end{figure}

Imposing BCs is an indispensable and extensively studied procedure in traditional solvers, from which we may get inspiration for neural operators. 
Taking the finite difference method (FDM) as an example, sometimes the nodes involved in the scheme are outside the computational domain near the boundary \cite{K.W.1994}.
These points are called ghost or fictitious points. 
To close the equation near the boundary, either (i) provide extra conditions to uniquely determine the value at ghost points \cite{Zhao2009,AsifFarooq2013}, or (ii) adopt asymmetric schemes for the boundary nodes to avoid the need for ghost points \cite{Nagarajan2003,Kim2007a,Berland2007}. 
The latter approach is easier to execute, which makes it a common practice.
The boundary treatment in FEM is similar to asymmetric schemes of FDM, because the nodes on the boundary are only related to the nodes inside the computational domain through the global matrix and not to those outside the boundary.
Then, a natural idea is to find an analogous operation for neural operators to impose BCs, which may, however, involve modifying the operator itself. 
If the time-marching process of the pre-trained neural operator $\mathcal{G}_{\theta}$ is expressed as $\boldsymbol{u}_{t+\Delta t}=\mathcal{G}_{\theta}(\boldsymbol{u}_t)$, then the boundary requires to architect and train a new neural operator $\boldsymbol{u}_{t+\Delta t}=\mathcal{G}_{\text{BC}}(\boldsymbol{u}_t)$ to ensure the output $\boldsymbol{u}_{t+\Delta t}$ satisfies the boundary condition. 
However, it is infeasible to train neural operators for all possible BCs, which could largely undermine the essential advantages of the method, reusability and then efficiency.
This suggests the former approach is probably more suitable for neural operators, i.e., to figure out the value at ghost points. 
In fact, the ghost point is closely associated to what we referred to as the “corrosion of domain” in \cite{Li2022c}, which requires a larger domain than the desired computational domain as the input, see Fig. \ref{fig_ghost_point}.
Here the nodes on the extended domain can be seen as ghost points. 
So, the question turns to how to decide the unique values on the extended domain.
Drawing such analogy and comparison to traditional solvers provide a thought for developing the method for imposing boundary conditions.

This paper aims to build up a universal pattern for imposing boundary conditions in the prediction of pre-trained LNO, which gives pre-trained LNO the reusability in solving problems with different boundary conditions. 
We make full use of the phenomenon `corrosion of domain' in LNO, to embody the information of boundary conditions on the extended domain in each time-marching step. 
In this way, the imposition of boundary conditions turns to a specific question of determining the values on the extended domain. 
The rest of the paper is organized as follows. 
Section \ref{sec2} introduces the preliminaries of the study, including the LNO and the governing equations to be learned. 
Section \ref{sec3} first demonstrates the framework of imposing boundary conditions in the prediction process of pre-trained LNO, and then presents strategies of introducing different BCs. 
These strategies are validated and compared in Section \ref{sec4} by three numerical examples. 
Finally, conclusions are drawn in Section \ref{sec5}.

\section{Preliminaries: local neural operator and compressible fluid dynamics}
\label{sec2}

\subsection{Definiiton of local neural operator}
\label{sec2:1}
\begin{figure}[htbp]
  \centering
  \includegraphics[width=0.65\textwidth]{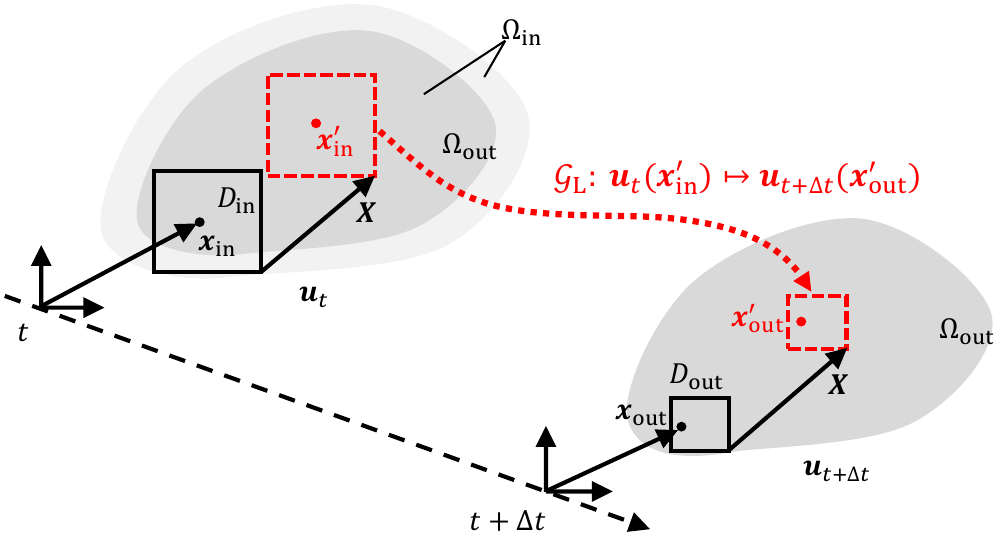}
  \caption{The time-marching local operator $\mathcal{G}_\text{L}$.
   $\boldsymbol{u}_t,\boldsymbol{u}_{t+\Delta t}$ are the physical fields at time level $t$ and $t+\Delta t$, respectively.
  $\Delta t$ is the time interval. 
  $\Omega_\text{out}/\Omega_\text{in}$ is the output/input computational domain.
  $D_\text{out}/D_\text{in}$ is the unit output/input domain.
  $\boldsymbol{X}$ is an arbitrary shifting vector.}\label{fig_LNOmapping}
\end{figure}

As shown in Fig. \ref{fig_LNOmapping}, the local neural operator (LNO) $\mathcal{G}_{\theta}$ \cite{Li2022c,Ye2023b} is defined to learn the following time-marching operator $\mathcal{G}_{\text{L}}$ that maps physical fields from one time level to the next:
\begin{equation}
  \begin{split}
    \mathcal{G}_\text{L}:\boldsymbol{u}_t\left(\boldsymbol{x}_\text{in}^\prime\right)\mapsto \boldsymbol{u}_{t+\Delta t}\left(\boldsymbol{x}_\text{out}^\prime\right),\quad t\geq 0,\boldsymbol{x}_\text{out}^\prime\in\Omega_{\text{out}},\boldsymbol{x}_\text{in}^\prime\in\Omega_{\text{in}},\\
    \Omega_\text{out}=\left\{\boldsymbol{x}_\text{out}^\prime=\boldsymbol{x}_\text{out}+\boldsymbol{X}\ \middle|\ \boldsymbol{x}_\text{out}\in D_\text{out},\boldsymbol{X}\in\mathcal{X}\right\}, \\
    \Omega_\text{in}=\left\{\boldsymbol{x}_\text{in}^\prime=\boldsymbol{x}_\text{in}+\boldsymbol{X}\ \middle|\ \boldsymbol{x}_\text{in}\in D_\text{in},\boldsymbol{X}\in\mathcal{X}\right\},
  \end{split}
  \label{eq:2:lno_definition}
\end{equation}
where $D_\text{out}$ is a designated unit output domain, and $D_\text{in}$ is determined by a local-related condition as
\begin{equation}
  \frac{\partial \boldsymbol{u}_{t+\Delta t}(\boldsymbol{x}_\text{out})}{\partial \boldsymbol{u}_t(\boldsymbol{x}_\text{in})}=0,\quad\forall \boldsymbol{x}_\text{out} \in D_\text{out},\boldsymbol{x}_\text{in}\notin D_\text{in}.
  \label{eq:2:local_related_condition}
\end{equation}
Here $\boldsymbol{u}_t,\boldsymbol{u}_{t+\Delta t}\in \mathbb{R}^{d_u}$ are the physical fields at time level $t$ and $t+\Delta t$, respectively.
$\Delta t$ is the time interval. 
$d_u$ is the number of physical fields. 
$\Omega_\text{out}/\Omega_\text{in}\subset\mathbb{R}^d$ is the output/input computational domain determined by the unit output/input domain $D_\text{out}/D_\text{in}$ and a set of shifting vectors $\mathcal{X}$. 
$d$ is the number of spatial dimensions.

The definition Eq. (\ref{eq:2:lno_definition}) is built upon two basic assumptions: the time marching operator $\mathcal{G}_\text{L}$ should be local-related and shift-invariant.
The local-related assumption says the output $\boldsymbol{u}_{t+\Delta t}$ at a point should be only related to input values $\boldsymbol{u}_{t}$ around this point if the time interval $\Delta t$ is finite, which constraints the size of $D_\text{in}$ in Eq. (\ref{eq:2:local_related_condition}). 
The shift-invariant assumption means $\mathcal{G}_\text{L}$ should be objective wherever the coordinate is, which reflects in the definition of $\Omega_\text{out}$ and $\Omega_\text{in}$ that the unit output/input domain $D_\text{out}/D_\text{in}$ can translate along an arbitrary shifting vector $\boldsymbol{X}$.
The unit output/input domains are, figuratively speaking, a pair of brushes that can move freely by $\boldsymbol{X}$ in the infinite space to draw any computational domain as wish.
With the two assumptions, the operator defined by Eq. (\ref{eq:2:lno_definition}) essentially accepts functions on different computational domains as the output/input, i.e., the reusablity on different computational domains is achieved.

To realize the LNO concept, the designed network architecture must satisfy the two assumptions.
Here we adopt the architecture from Ref. \cite{Ye2023b}, which mainly comprised of the spectral path, physical path, lifting layer, and projection layer, which is briefly introduced in \ref{app:architecture}.
The architecture includes $4$ hyper-parameters, including the number of blocks $n$, the order of spectral transform $N$, the number controlling the stride of the spectral transform $K$, and the order of low-pass filtering in the spectral path $M$.
We set $n=4,N=12,K=2,M=6$, which is the best choice for the current learning problem according to \cite{Ye2023b}.

A consequence of the two assumptions is the output domain $\Omega_\text{out}$ is always smaller than the input domain $\Omega_\text{in}$, the reduced region is marked by the light gray color in Fig. \ref{fig_LNOmapping}.
This phenomenon is named 'corrosion of domain', which brings additional challenges for the application of LNO compared to other neural operators, but also offers an opportunity to impose BCs, which will be studied in the rest of the paper.
The width of the corroded domain depends on the architecture of LNO.
For the architecture adopted in this paper, the unit output domain is withsize $\frac{N}{K}=6$, the unit input domain is $\frac{N}{K}+2\frac{K-1}{K}nN=54$, the corroded width is $\frac{K-1}{K}nN=24$.

\subsection{Learning task: compressible fluid dynamics}
\label{sec2:2}
In this work, LNO is applied to learn and predict compressible viscous  flows. 
The governing equations of the nondimensional 2-D compressible viscous flow are
\begin{equation}
  \frac{\partial \rho}{\partial t}+\nabla \cdot(\rho \boldsymbol{v})=0,
 \label{eq:2:governing_equation1}
\end{equation}
\begin{equation}
\frac{\partial \rho \boldsymbol{v}}{\partial t}+\nabla \cdot(\rho \boldsymbol{v} \boldsymbol{v})=-\nabla p+\nabla \cdot \boldsymbol{\tau},
\label{eq:2:governing_equation2}
\end{equation}
\begin{equation}
\frac{\partial \rho E}{\partial t}+\nabla \cdot[(\rho E+p) \boldsymbol{v}]=\nabla \cdot(\boldsymbol{\tau} \cdot \boldsymbol{v})+\nabla \cdot \boldsymbol{q},
\label{eq:2:governing_equation3}
\end{equation}
with
\begin{equation}
  p=\frac{1}{\gamma\textit{Ma}^2}\rho T,
  \label{eq:2:ideal_gas_state}
\end{equation}
\begin{equation}
  \boldsymbol{\tau}=-\frac{2}{3\textit{Re}}(\nabla\cdot\boldsymbol{v})\boldsymbol{I}+\frac{2}{\textit{Re}}(\nabla\boldsymbol{v}+\nabla\boldsymbol{v}^T),
  \label{eq:2:viscous_stress_tensor}
\end{equation}
\begin{equation}
  \boldsymbol{q}=\frac{1}{(\gamma-1)\textit{Pr}\cdot\textit{Re}\cdot\textit{Ma}^2} \nabla T,
  \label{eq:2:viscous_heat_transfer}
\end{equation}
\begin{equation}
  E=\frac{|\boldsymbol{v}|^2}{2}+\frac{1}{(\gamma-1)\gamma\textit{Ma}^2} T.
  \label{eq:2:total_energy}
\end{equation}
Here $\rho,\boldsymbol{v},T,p,\boldsymbol{\tau},\boldsymbol{q},E$ are respectively the nondimensional density, velocity, temperature, pressure, viscous stress, heat conduction term, and total energy.
There are four dimensionless numbers describing the property of fluid, including specific heat ratio $\gamma$, Reynolds number $Re$, Mach number $Ma$, and Prandtl number $Pr$. 
In this work, the four numbers are fixed as $\gamma=1.4,Re=100,Ma=0.2,Pr=0.72$. 
$\rho,\boldsymbol{v},T$ are chosen as the independent variables to be solved. 
We denote $\boldsymbol{u}=\{\rho,T,v_x,v_y\}$ and abbreviate $\boldsymbol{u}(\boldsymbol{x},t)$ as $\boldsymbol{u}_t$.

The governing equation Eqs. (\ref{eq:2:governing_equation1}-\ref{eq:2:governing_equation3}) must cooperate with specific conditions (including IC, BC, and the computational domain) to obtain the solution of a certain flow problem.
Denote the computational domain as $\Omega$, the IC and BC are formulated as:
\begin{equation}
  \begin{split}
  \text{Initial condition:}\boldsymbol{u}(\boldsymbol{x},0)=\boldsymbol{u}_0(\boldsymbol{x}),\quad \boldsymbol{x}\in\Omega,\\
  \text{Boundary condition:}f(\boldsymbol{u}(\boldsymbol{x},t))=0,\quad \boldsymbol{x}\in \partial \Omega,t\geq 0.
  \end{split}
  \label{eq:2:BC_IC}
\end{equation}
The formulation of $f$ can be multifarious, such as the no-slip velocity BC $\boldsymbol{v}=\text{const}$, the heat flux BC $\frac{\partial T}{\partial x}=\text{const}$, the outflow BC, far-field BC, and so on.

\subsection{Train LNO to learn compressible fluid dynamics}
\label{sec2:3}
This subsection briefly illustrates the key elements of training LNO. 
All the procedures and settings are the same as Ref. \cite{Ye2023b}. 

To generate abundant data samples, we choose a flow problem with random initial condition
\begin{equation}
  \boldsymbol{u}_0=\left \{ 
  \begin{matrix}
      1\\
      1\\
      0.6[\sin\pi x,\sin 2\pi x,\cos \pi x,\cos 2\pi x]\boldsymbol{\Lambda}_1[\sin \pi y,\sin 2\pi y,\cos \pi y,\cos 2\pi y]^\text{T}\\
      0.6[\sin\pi x,\sin 2\pi x,\cos \pi x,\cos 2\pi x]\boldsymbol{\Lambda}_2[\sin \pi y,\sin 2\pi y,\cos \pi y,\cos 2\pi y]^\text{T}
  \end{matrix}
   \right \}
   \label{eq:3:random_initial_condition}
\end{equation}
in a square domain $\left[-1,1\right]\times[-1,1]$ with periodic boundary conditions around. Here $\boldsymbol{\Lambda}_k=\left\{\lambda_{k,ij}\right\}\ (i,j=1~4,k=1~2),\ \ \lambda_{k,ij}\sim N(0,1)$. 
Data samples are numerically calculated from the governing equation (Eqs. (\ref{eq:2:governing_equation1}-\ref{eq:2:governing_equation3})) and initial condition Eq. (\ref{eq:3:random_initial_condition}) with linear FEM and the fourth-order Runge-Kutta scheme. 
The current LNO architecture requires input and output on equidistant nodes, so the square domain is discretized to an equidistant cartesian mesh with $\Delta x=1/64$. 
The flow field is recorded with time interval $\Delta t=0.05$ until $t=5$ as a piece of data samples. 
Totally $225$ pieces of samples from different initial conditions are recorded, where $200$ of them are used as training data and the left $25$ are used for validation.

\begin{figure}[htbp]
  \centering
  \includegraphics[width=0.9\textwidth]{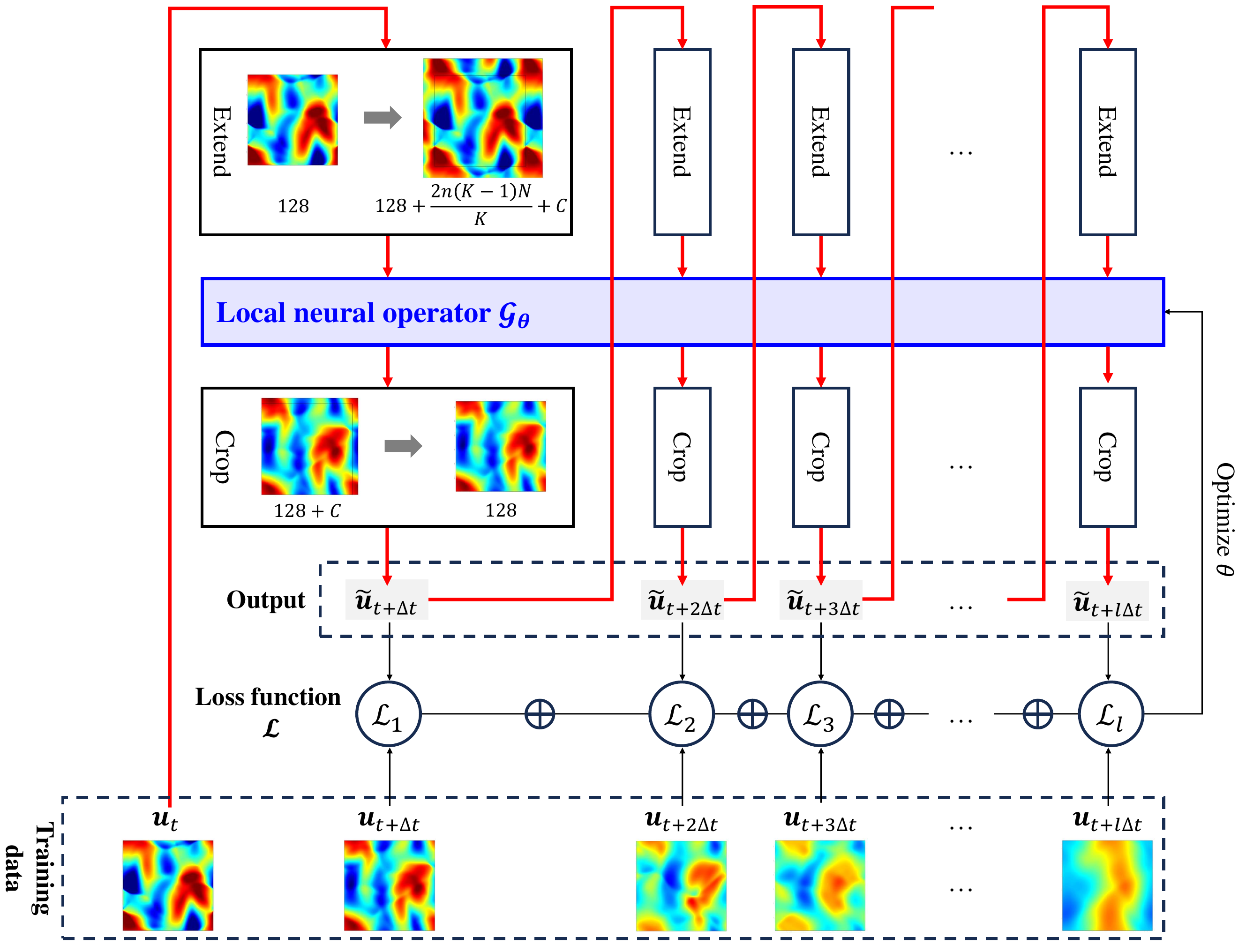}
  \caption{The training process of LNO. $C$ is the extra extended size to ensure the input is exactly covered by an integer number of unit input domain.}\label{fig_trainingprocess}
\end{figure}

One training iteration of LNO is depicted in Fig. \ref{fig_trainingprocess}. 
A successive series of samples $\{\boldsymbol{u}_{t+i\Delta t}|i=0,1,2,\ldots,l\}$ is extracted from the training dataset, where $t$ is a random moment range from $0$ to $5-l\Delta t$.
$l$ is the number of continuous prediction steps, which is set as 10. 
$\boldsymbol{u}_t$ is the first input sent to LNO to predict the flow field at $t+\Delta t$. 
Denote the output of LNO as $\tilde{\boldsymbol{u}}_{t+\Delta t}$. 
Before/after LNO prediction, the flow field is extended/cropped to tackle the corrosion of domain and make sure the final output is with the same size as the input. 
The size of extension is $\frac{K-1}{K}nN=24$ on both sides along with an additional $C=4$ on one side to ensure the input size is divisible by $\frac{N}{K}$.
In the cropping process, the additional size $C$ is removed to recover to the original size.
Then $\tilde{\boldsymbol{u}}_{t+\Delta t}$ is used as the next input to LNO again to predict the solution at the subsequent moment. 
LNO totally predicts for $l$ times to give $l$ predicted flow fields $\{\tilde{\boldsymbol{u}}_{t+i\Delta t}| i=1,2,\ldots,l\}$. 
The loss function $\mathcal{L}$ is defined as the difference between the predicted solutions and data samples:
\begin{equation}
  \mathcal{L}=\frac{1}{l}\sum_{i=1}^{l}\|\boldsymbol{u}_{t+i\Delta t}-\tilde{\boldsymbol{u}}_{t+i\Delta t}\|_2,
  \label{eq:2:loss_function}
\end{equation}
where
\begin{equation}
  \tilde{\boldsymbol{u}}_{t+i\Delta t}=\mathcal{G}_{\theta}\left (\tilde{\boldsymbol{u}}_{t+(i-1)\Delta t}\right ),\quad i\geq 1,
\end{equation}
in which $\tilde{\boldsymbol{u}}_{t}=\boldsymbol{u}_{t}$.
The value of $\mathcal{L}$ reflects the quality of LNO prediction.
The smaller $\mathcal{L}$ is, the more accurate the predicted flow field is.
LNO predicts for $l$ successive steps instead of a single step for numerical stability. 
Applications usually require observing the dynamic evolution or predicting the flow field after a long period, which requires LNO to be numerically stable in the long-term time-marching process. 
The loss function defined in Eq. (\ref{eq:2:loss_function}) makes sure the prediction error does not blow up after $l$ steps, which enhances the stability of LNO.

Then, backpropagate from $\mathcal{L}$ to optimize the weights $\theta$ in order to reduce $\mathcal{L}$.
The training process contains $200$ epochs with $500$ iterations in each epoch, i.e., totally ${10}^5$ iterations. 
The optimizer is Adam \cite{Kingma2015} with an initial learning rate $0.001$. 
The learning rate is manually multiplied by $0.7$ every $20$ epochs. 

After training, we obtain an LNO with optimized weights $\theta$ that can predict the fluid dynamics. 
In this paper a pre-trained LNO with hyper-parameters $n=4, N=12, M=6, K=2$ from Ref. \cite{Ye2023b} is adopted for further discussion and applications. 
The averaged prediction errors of this pre-trained model on the validation samples are $0.0650, 0.0475, 0.0445, 0.0518, 0.0694$ for velocity at $t=0.2,0.5,1,2,5$.

\section{Imposing boundary conditions in the prediction of pre-trained LNO}
\label{sec3}
Of the four elements that determine the solution of a concrete problem, i.e.,the governing equations and problem-specific conditions including IC, BC, and the computational domain, one pre-trained LNO covers the reusability of three except for the BC. 
The physical law behind the governing equations is learned by LNO from training samples; 
accepting different ICs as the input function is an essential feature of the time-marching neural operators as an operator does not restrict the value of input functions; 
the universality of various computational domains can be realized by shifting the unit input and output domain according to the shift-invariant assumption. 
However, the pre-trained LNO itself cannot handle different boundary conditions. 
On the one hand, the motion of fluids near the boundary differs from the interior region due to the influence of the boundary. 
Thus, one pre-trained LNO cannot predict the flow in both regions. 
On the other hand, due to ‘corrosion of domain’, LNO cannot output the prediction on the corroded domain.
The two facts suggest that boundary treatments are indispensable for LNO to complete the reusability in all four elements and deal with the corroded domain. 

The above two facts are connected intrinsically. 
According to Ref. \cite{Ye2023b}, the corroded domain must fully cover the region affected by boundaries within one time step to guarantee good prediction of LNO. 
So, in the following practice with a well-trained LNO, we mostly talk about the corroded domain rather than the hard-to-tell boundary-affected region.

\begin{figure}[htbp]
  \centering
  \includegraphics[width=0.9\textwidth]{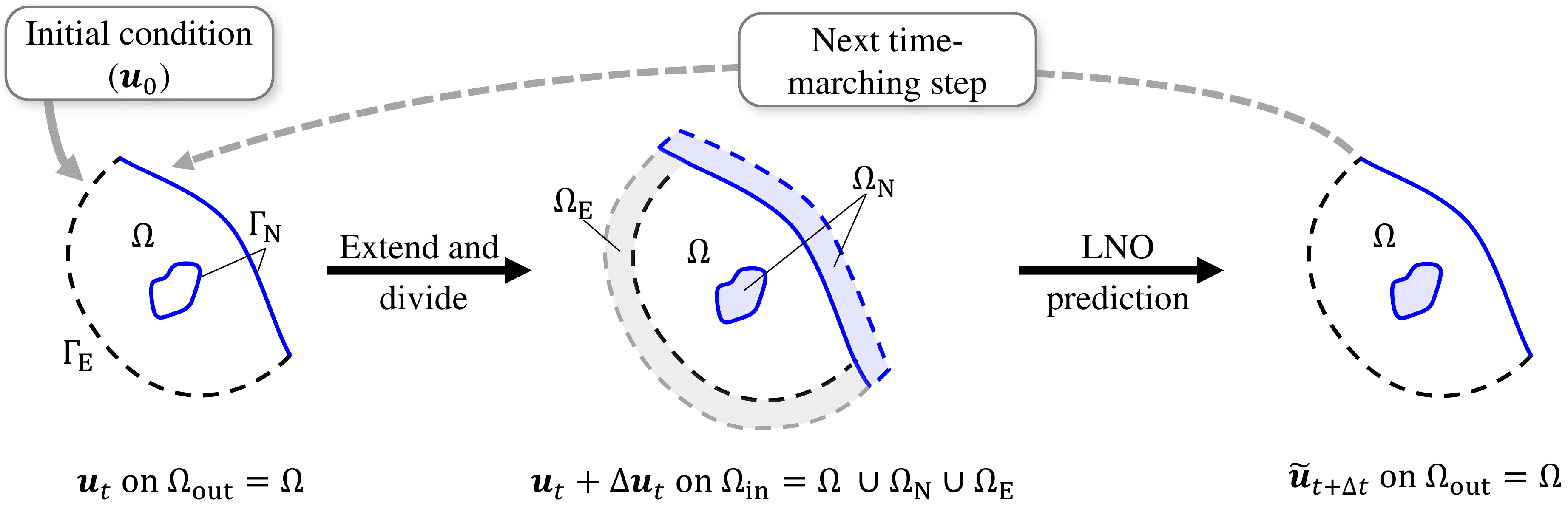}
  \caption{Time-marching prediction process of pre-trained LNO with boundary treatment}\label{fig_vde_process}
\end{figure}

To introduce the effect of the BC and prevent the corrosion of domain, a framework of applying pre-trained LNO to solve a flow problem is designed as shown in Fig. \ref{fig_vde_process}. 
The initial condition $\boldsymbol{u}_t$ on $\Omega$ is known. 
Before sending $\boldsymbol{u}_t$ to the pre-trained LNO, $\Omega$ is extended to a larger domain $\Omega_\text{in}$. 
The extension width is set as the width of corrosion ($\frac{K-1}{K}nN$ for the current LNO architecture). 
The function on $\Omega_\text{in}$ is used as the input of LNO to output the prediction on $\Omega$, so  $\Omega_\text{out}=\Omega$. 
To distinguish the known IC on $\Omega$ and the extended input function on $\Omega_\text{in}$,  we use $\boldsymbol{u}_t$ to refer specifically the given IC on $\Omega$ which equals to zero on the extended domain in the rest of the paper.
A new function $\Delta\boldsymbol{u}_t$ is introduced to represent all the changes to transform $\boldsymbol{u}_t$ on $\Omega$ to the extended input function on $\Omega_\text{in}$.
The changes include assigning values for $\rho,T,v_x,v_y$ on the extended domain and possible corrections made on $\Omega$.
In this way, the extended input function on $\Omega_\text{in}$ is forumulated as $\boldsymbol{u}_t+\Delta\boldsymbol{u}_t$, which is sent to LNO to output $\tilde{\boldsymbol{u}}_{t+\Delta t}$. 
After the corrosion in LNO prediction, the output domain is the same as $\Omega$. 
The process is formulated as
\begin{equation}
  \tilde{\boldsymbol{u}}_{t+\Delta t}(\boldsymbol{x}_\text{out})=\mathcal{G}_{\theta}(\boldsymbol{u}_t+\Delta \boldsymbol{u}_t)(\boldsymbol{x}_\text{in}),\quad \boldsymbol{x}_\text{in}\in\Omega_\text{in},\boldsymbol{x}_\text{out}\in\Omega=\Omega_\text{out},
  \label{eq:3:vde_formulation}
\end{equation}
where $\mathcal{G}_\theta$ is the pre-trained LNO with fixed learned weights $\theta$.
Then, $\tilde{\boldsymbol{u}}_{t+\Delta t}$ on $\Omega$ is used as the IC of the next time-marching step. 
In this way, the computational domain remains unchanged in long-term prediction.

The prediction requires the value of $\Delta\boldsymbol{u}_t$, which should be given according to the type of BC to make the output value on the boundary curve close to BC. 
Therefore, the focus of the boundary treatment is to decide the assigned value $\Delta\boldsymbol{u}_t$ on the extended domain. 
We collectively name the boundary treatments through assigning value on the extended domain as virtual domain extension (VDE).
We categorize the common BCs in computational fluid dynamics into two types, extendable BCs and non-extendabal BCs, and discuss their strategies to calculate $\Delta\boldsymbol{u}_t$ respectively in the following contexts. 

\subsection{Imposing extendable boundary conditions}
\label{sec3:1}
\begin{figure}[htbp]
  \centering
  \includegraphics[width=0.85\textwidth]{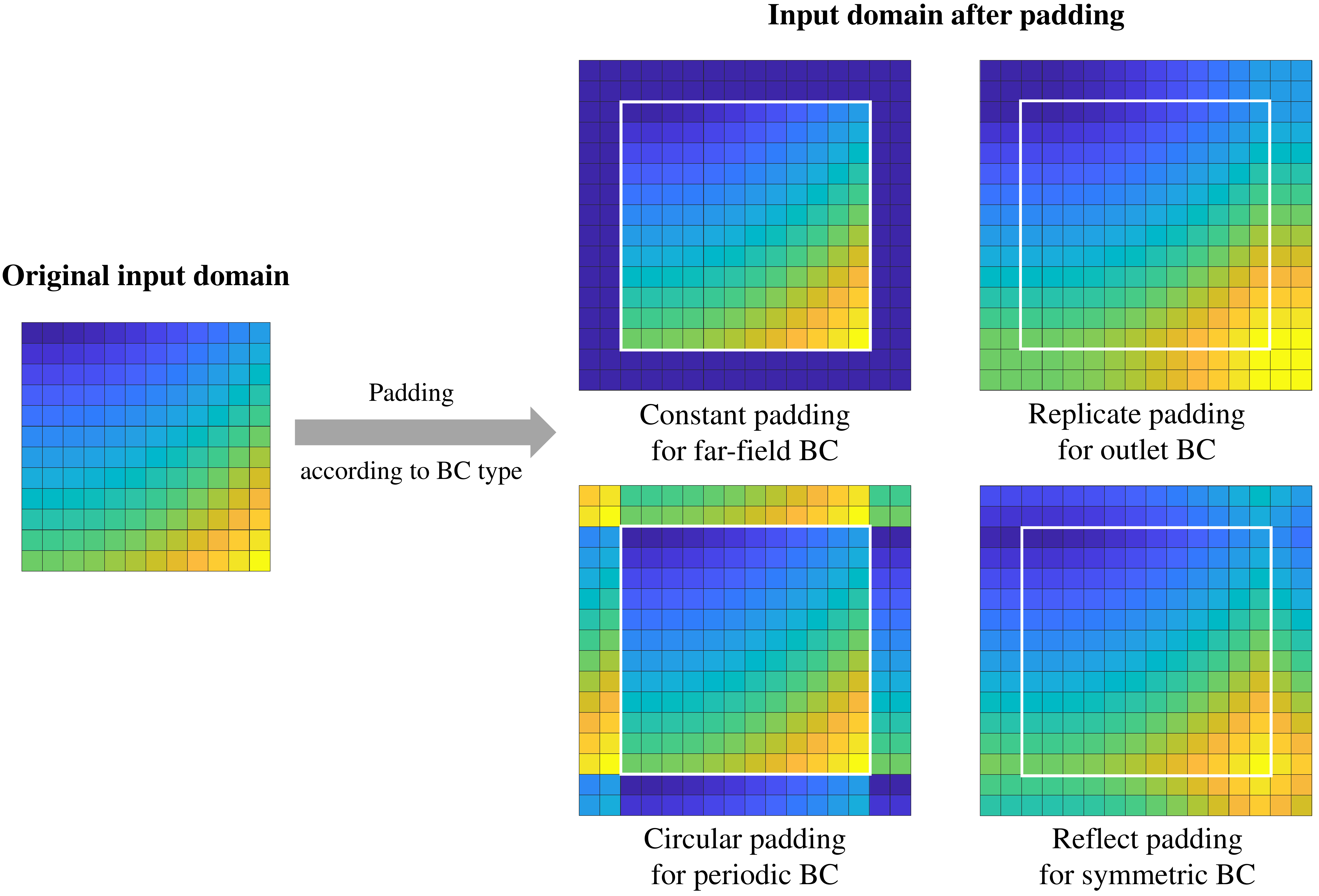}
  \caption{Schematic of the input computational domain before and after padding operation. Each colored block represents one node in the space. Here the extension width is set as $2$ just for demonstration purposes.}\label{fig_padding}
\end{figure}

The first category of BCs is extendable BCs, also called artificial BCs, denoted as $\Gamma_\text{E}$ and marked by black dotted line in Fig. \ref{fig_vde_process}. 
The domain extended from $\Gamma_\text{E}$ is denoted as $\Omega_\text{E}$. 
The extendable boundaries mean the computational domain is cutoff intentionally at the boundary by the user because the flow filed outside the boundary is known without complex computation, e.g., the periodic BC, symmetric BC, far-field BC, and the outflow BC. 
These BCs are easy to handle in LNO prediction by assigning the known flow field on the extended domain with the padding operation. 
Obviously, after padding operation the BC is strictly satisfied on $\Gamma_\text{E}$. 
The deep learning toolkit PyTorch provides several handy padding options, which cover the most of the extendable BCs. 
The options for different BCs are listed in Table \ref{tab_padding_option}, and the schematic of flow field before and after padding is in Fig. \ref{fig_padding}. 
In fact, padding operation is already incorporated in the training process of LNO.
As illustrated in Fig. \ref{fig_trainingprocess}, the input is extended before being sent into LNO.
Concretely, circular padding is used here, because the training samples all have periodic BCs.

\begin{table}[htbp]
  \centering
  \caption{Padding option for different types of extendable boundary conditions}\label{tab_padding_option}
  \begin{tabular}{cc}
    \toprule
    Extendable BC & Padding option\\
    \midrule
    Far-field & Padding=`constant’\\
    Outflow & Padding=`replicate’\\
    Periodic & Padding=`circular’\\
    Symmetric & Padding=`reflect’\\
    \bottomrule
  \end{tabular}
  
\end{table}

\subsection{Imposing non-extendable boundary conditions}
\label{sec3:2}
In contrast to the extendable BCs, the second type is non-extendable ones $\Gamma_\text{N}$ marked by blue solid line in Fig. \ref{fig_vde_process}, such as no-slip BC, heat flux BC, etc. 
The domain extended from $\Gamma_\text{N}$ is denoted as $\Omega_\text{N}$.
Non-extendable BCs can be further divided into external (such as the solid wall for flow in a tube) and internal ones (such as the solid wall in the flow over a circular cylinder), which are both illustrated in Fig. \ref{fig_vde_process}. 
The main difference is that for the internal BCs, the domain outside the boundary is an enclosed domain that would not be corroded in LNO prediction, but it still needs assigned value to impose the boundary condition. 
The non-extendable BCs may be on complex curvilinear geometries, and the extended domain outside the BCs contains either unknown flows or solid regions.
These chanllenges make the boundary treatment for non-extendable BCs the major topic of this paper. 

This subsection discusses the strategies to calculate $\Delta\boldsymbol{u}_t$ for non-extendable boundary conditions.
We begin with borrowing ideas from the boundary treatment in traditional solvers, then improve it according to the characteristics of LNO.
The following discussion uses the no-slip stationary wall
\begin{equation}
  \boldsymbol{v}(\boldsymbol{x},t)=\boldsymbol{0},\quad \boldsymbol{x}\in\Gamma_\text{N}
  \label{eq:3:noslip_wall}
\end{equation}
as the example for convenience; nonetheless, most of the strategies can be generalized to other non-extendable BCs.


\subsubsection{Direct imposing like traditional solvers}
\label{sec3:2:1}
\begin{figure}[htbp]
  \centering
  \includegraphics[width=\textwidth]{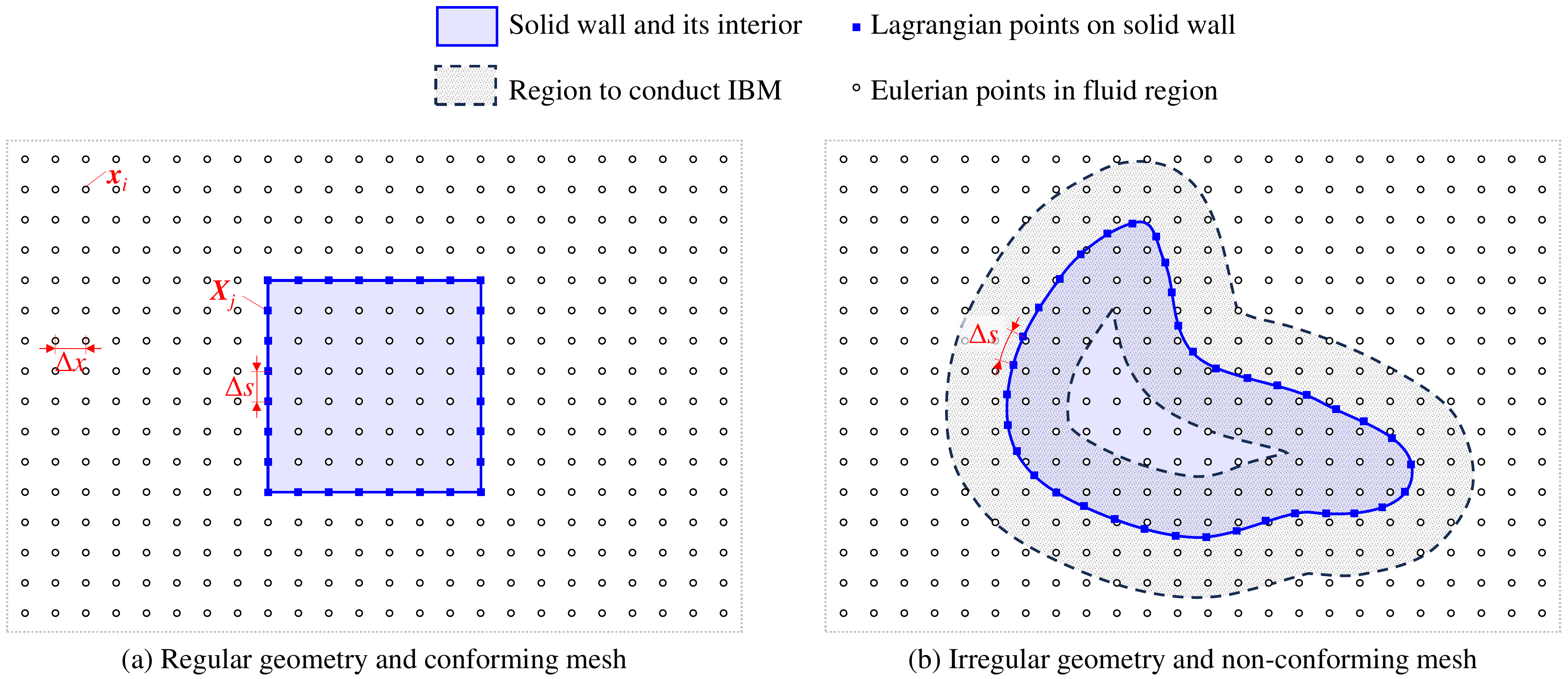}
  \caption{Schematic of direct imposing}\label{fig_direct_imposing}
\end{figure}

To deal with non-extendable BCs, the first thought comes to mind is to borrow ideas from traditional numerical solvers. 
This is what we have done in Ref. \cite{Li2022c}.
When solving the governing equation Eqs. (\ref{eq:2:governing_equation1}-\ref{eq:2:governing_equation3}) with an explicit temporal scheme (such as forward Euler or explicit Runge-Kutta), the boundary condition Eq. (\ref{eq:3:noslip_wall}) is commonly imposed by directly altering the nodal value on $\Gamma_\text{N}$ at each time level.
The boundary nodes are as marked by the blue solid square in Fig. \ref{fig_direct_imposing}(a). 
In the case the nodes are not located at the boundary curve (i.e., non-conforming mesh) as Fig. \ref{fig_direct_imposing}(b), researchers developed the immersed boundary method (IBM) to turn the influence of the boundary into an equivalent body force $\boldsymbol{f}_\text{N}$ on the nodes around the boundary to correct on the solution:
\begin{equation}
  \boldsymbol{u}_t=\boldsymbol{u}_t^\ast+\boldsymbol{f}_\text{N}\Delta t,
  \label{eq:3:IBM_start}
\end{equation}
where $\boldsymbol{u}_t^\ast$ is the intermediate flow field without considering the boundary. 
$\boldsymbol{f}_\text{N}$ is non-zero within a small distance from $\Gamma_\text{N}$, which is marked gray in Fig. \ref{fig_direct_imposing}(b).

$\boldsymbol{f}_\text{N}$ is calculated as followed. 
First interpolate the intermediate flow field $\boldsymbol{u}_t^\ast$ on the equidistant computational grids $\boldsymbol{x}_i$ (also called Eulerian points) to points $\boldsymbol{X}_j$ on the boundary curve (also called Lagrangian points):
\begin{equation}
  \boldsymbol{U}_t^{*}(\boldsymbol{X}_j)=\sum_{i=1}^{N_\text{G}^{\text{Euler}}}\boldsymbol{u}_t^{*}(\boldsymbol{x}_i)\delta_h(\boldsymbol{x}_i-\boldsymbol{X}_j)\Delta x^2,
  \label{eq:3:IBM_begin}
\end{equation}
where the uppercase letters $\boldsymbol{U}$,$\boldsymbol{X}$ denote variables on Lagrangian points, and lowercase letters $\boldsymbol{u},\boldsymbol{x}$ denote variables on Eulerian points. 
$N_\text{G}^{\text{Euler}},N_\text{G}^{\text{Lagrange}}$ is the number of Eulerian and Lagrangian points, respectively.
$\delta_h$ is the approximation for the impulse function $\delta$, here the 2-node piecewise function is chosen for simplicity:
\begin{equation}
  \delta_h(\boldsymbol{x}-\boldsymbol{X})=\frac{1}{\Delta x^2}w\left (\frac{x-X}{\Delta x}\right )w\left (\frac{y-Y}{\Delta x}\right )
\end{equation}
\begin{equation}
  w(x)=\left\{\begin{array}{c}
      1-|x|,|x|<1,\\
      0,|x| \geq 1 .
      \end{array}\right.
\end{equation}
Then, calculate the equivalent body force $\boldsymbol{F}_\text{N}$ on Lagrangian points to enforce the boundary condition $\boldsymbol{U}=\boldsymbol{u}_\text{BC}$:
\begin{equation}
  \boldsymbol{F}_\text{N}=\frac{\boldsymbol{u}_\text{BC}-\boldsymbol{U}_t^\ast}{\Delta t}.
\end{equation}
Finally, interpolate $\boldsymbol{F}_\text{N}$ to Eulerian points to obtain $\boldsymbol{f}_\text{N}$:
\begin{equation}
  \boldsymbol{f}_\text{N}(\boldsymbol{x}_i)=\sum_{j=1}^{N_\text{G}^{\text{Lagrange}}}\boldsymbol{F}_\text{N}(\boldsymbol{X}_j)\delta_h(\boldsymbol{x}_i-\boldsymbol{X}_j)\Delta x\Delta s,
\end{equation}
where $\boldsymbol{f}_\text{N},\boldsymbol{F}_\text{N}$ are the equivalent body force on Eulerian and Lagrangian points, respectively;
$\Delta s$ is the distance between Lagrangian points as marked in Fig. \ref{fig_direct_imposing}.
As $\delta_h$ is a 2-node piecewise function, $\boldsymbol{f}_\text{N}$ is non-zero only within $\Delta x$ from $\Gamma_\text{N}$.

This boundary treatment can be applied to the LNO prediction with a few adjustments. 
Firstly, in traditional solvers the BC is imposed \textit{after} the time-marching, whereas the boundary treatment should be \textit{before} time-marching in LNO prediction as shown in Fig. \ref{fig_vde_process}. 
Thus, the boundary treatment is moved to be implemented on the extended input function. 
Secondly, this boundary treatment should be made on a known flow field, which does not exist on the extended domain $\Omega_\text{N}$.
We artificially set the extended input function on $\Omega_\text{N}$ as the boudnary value $\boldsymbol{u}_{\text{BC}}$ to allow the following boundary treatment, i.e., the intermediate flow field $\boldsymbol{u}_t^\ast$ in Eq. (\ref{eq:3:IBM_start}) to be corrected is $\boldsymbol{u}_t+h\boldsymbol{u}_{\text{BC}}$ on $\Omega_\text{in}$.
Here $h$ is an indicator of whether $x\in\Omega_\text{N}$:
\begin{equation}
  h(x)=\left\{\begin{array}{c}
    1,\quad x\in\Omega_\text{N},\\
    0,\quad \text{otherwise}.
      \end{array}\right.
      \label{eq:3:location_indicator}
\end{equation}

In this way, the direct imposition of BCs in LNO prediction is formulated as
\begin{equation}
  \tilde{\boldsymbol{u}}_{t+\Delta t}(\boldsymbol{x}_\text{out})=\mathcal{G}_\theta(\boldsymbol{u}_t+h\boldsymbol{u}_{\text{BC}}+\boldsymbol{f}_\text{N}\Delta t)(\boldsymbol{x}_\text{in}),
  \label{eq:3:direct_imposing1}
\end{equation}
i.e.,
\begin{equation}
  \Delta \boldsymbol{u}_t=h\boldsymbol{u}_{\text{BC}}+\boldsymbol{f}_\text{N}\Delta t.
\end{equation}
We name this strategy the `direct imposition'.

One exception is for interior BCs, $\boldsymbol{u}_t$ has values on $\Omega_\text{N}$ (the enclosed domain by the boundary curve), so assigning $\boldsymbol{u}_{\text{BC}}$ on $\Omega_\text{N}$ becomes optional.
One can choose to operate the boundary treatment as Eq. (\ref{eq:3:direct_imposing1}) or simply make correction on the original values of $\boldsymbol{u}_t$:
\begin{equation}
  \tilde{\boldsymbol{u}}_{t+\Delta t}(\boldsymbol{x}_\text{out})=\mathcal{G}_\theta(\boldsymbol{u}_t+\boldsymbol{f}_\text{N}\Delta t)(\boldsymbol{x}_\text{in}).
  \label{eq:3:direct_imposing_curve}
\end{equation}
Eq. (\ref{eq:3:direct_imposing_curve}) only makes correction on the region within $\Delta x$ from $\Gamma_\text{N}$.
To distinguish the above two options, the strategy in Eq. (\ref{eq:3:direct_imposing_curve}) is named `direct imposition on boundary curve'.
The two options are both available for traditional solvers and provide nearly the same results. 
In Ref. \cite{Wu2009} where the latter option is used, an isolated flow is generated inside the enclosed domain, which does not affect the flow outside. 
However, the two options differ for LNO due to its different feature from traditional solvers, which will be demonstrated by the numerical examples in Section \ref{sec4}.

\subsubsection{Imposing with pressure symmetry}
\label{sec3:2:2}
The solid wall boundary condition usually offers explicit formulations for $\boldsymbol{v}$ and $T$, but not $\rho$. 
But one thing special in LNO prediction is that the extended domain requires the definite value of every input variable including $\rho$. 
Therefore, it is crucial to assign a reasonable value to $\rho$ to ensure accurate imposing of BC. 
This subsection provides an approximate and easy rule to compute $\rho$ on the extended domain based on pressure symmetry.


Ignore the effect of viscosity, the 2-D momentum equation for the inviscid flow is:
\begin{equation}
  \rho \frac{\partial v_x}{\partial t}+v_x\frac{\partial v_x}{\partial x}+v_y\frac{\partial v_x}{\partial y}=-\frac{\partial p}{\partial x},
  \label{eq:3:inviscid_governing1}
\end{equation}
\begin{equation}
  \rho \frac{\partial v_y}{\partial t}+v_x\frac{\partial v_y}{\partial x}+v_y\frac{\partial v_y}{\partial y}=-\frac{\partial p}{\partial y}.
  \label{eq:3:inviscid_governing2}
\end{equation}
Eq. (\ref{eq:3:inviscid_governing1})$\times n_x+$Eq. (\ref{eq:3:inviscid_governing2})$\times n_y$ and ignore the change of $n_x$,$n_y$ in space:
\begin{equation}
  \rho\frac{\partial (v_xn_x+v_yn_y)}{\partial t}+v_x\frac{\partial (v_xn_x+v_yn_y)}{\partial x}+v_y\frac{\partial (v_xn_x+v_yn_y)}{\partial x}\approx -\left (n_x\frac{\partial p}{\partial y}+n_y\frac{\partial p}{\partial y}\right ).
  \label{eq:3:PS_derive1}
\end{equation}
The solid wall BC requires the impermeability along the normal direction:
\begin{equation}
  \boldsymbol{v}\cdot \boldsymbol{n}=v_x n_x+v_y n_y=0.
  \label{eq:3:non_permeable_BC}
\end{equation}
Substitute Eq. (\ref{eq:3:non_permeable_BC}) to Eq. (\ref{eq:3:PS_derive1}), the left-hand side of Eq. (\ref{eq:3:PS_derive1}) equals zero, so the right-hand side is also zero:
\begin{equation}
  \left (n_x\frac{\partial p}{\partial y}+n_y\frac{\partial p}{\partial y}\right )=\frac{\partial p}{\partial n}=0,
  \label{eq:3:pressure_symmetry}
\end{equation}
which shows the pressure should be symmetric about the solid wall. 
The pressure gradient is one of the sources driving the flow, so setting the pressure gradient along the normal direction to zero effectively reduces the non-physical flow through the solid wall.

\begin{figure}[htbp]
  \centering
  \includegraphics[width=0.85\textwidth]{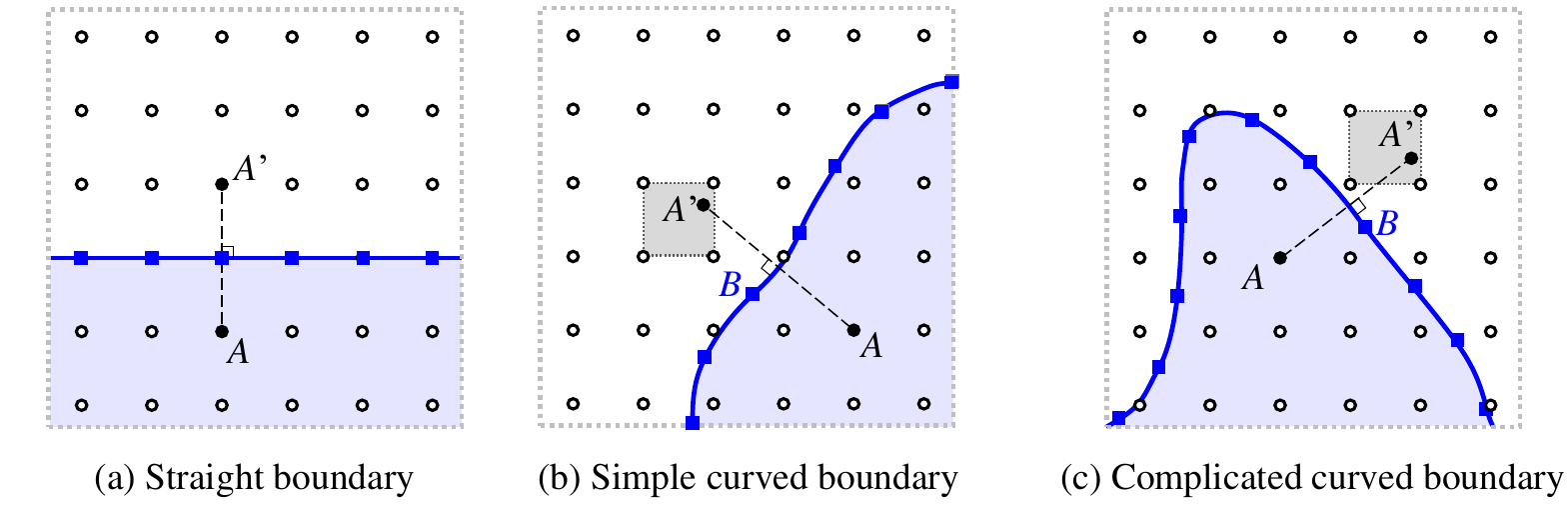}
  \caption{The symmetric point $A^\prime$ of a node $A$ inside the solid wall}\label{fig_pressure_symmetry}
\end{figure}

With Eq. (\ref{eq:3:pressure_symmetry}), $\rho$ on the extended domain is calculated easily as follows.
Denote a node on $\Omega_\text{N}$ as point $A$, the symmetric point about the boundary is $A^\prime$, as shown in Fig. \ref{fig_pressure_symmetry}(a). 
According to Eq. (\ref{eq:3:pressure_symmetry}), the pressure at $A$ and $A^\prime$ should be the same. 
With the equation of state $p=\rho RT$, the formulation for the density $\rho_A$ is derived:
\begin{equation}
  \rho_A R T_A=\rho_{A^{\prime}} R T_{A^{\prime}},
\end{equation}
\begin{equation}
  \rho_A=\frac{\rho_{A^{\prime}} T_{A^{\prime}}}{T_A}.
  \label{eq:3:density_symmetry}
\end{equation}
This leads to
\begin{equation}
  \Delta \rho_A=h\rho_A=h\frac{\rho_{A^\prime}T_{A^\prime}}{\Delta T_A},
\end{equation}
where $\Delta T_A$ is the temperature assigned on $\Omega_\text{N}$ computed by direct imposition or other strategies,
$h$ is to emphasize pressure symmetry is implemented only for $\Omega_\text{N}$ to avoid confusion in notation.

If the symmetric point $A^\prime$ is not a node on the Cartesian mesh, $\rho_{A^\prime}$ and $T_{A^\prime}$ are interpolated from the closest four points, see Fig. \ref{fig_pressure_symmetry}(b). 
Furthermore, if the boundary curve is complicated as shown in Fig. \ref{fig_pressure_symmetry}(c), the symmetric point is uniquely decided by the algorithm from Ref. \cite{Mittal2008}: 
first, calculate the distance from $A$ to all Lagrangian points (the blue squares) to find the closest one $B$; 
second, calculate the distance from $A$ to two boundary line segments with $B$ as the endpoint; 
third, find the intersection of the closer segment with the perpendicular line from $A$. 
If the intersection is on the segment, it is the center of symmetry. 
If not, use the intersection with the other segment. 
If the second intersection is not on the segment as well, use the Lagrangian point $B$ as the center of symmetry. 
Then, the symmetric point $A^\prime$ is determined. 
This algorithm makes sure a unique symmetric point for every node on the extended domain.

To assign values for $\rho,T,\boldsymbol{v}$ on the extended domain, direct imposition and pressure symmetry should be both adopted, with direct imposition to calculate $T,\boldsymbol{v}$ and pressure symmetry to calculate $\rho$.
We refer this combination as pressure symmetry in the rest of this paper for simplicity.
Moreover, to illustrate the necessity of assigning density values, the density will be set as a constant number in the direct imposition strategy in the numerical examples.
All these settings and formulations are summarized in Table \ref{tab_algorithm_summary}.

\subsubsection{Imposing with backpropagation optimization}
\label{sec3:2:3}
The above two strategies both rely on the theory of fluid dynamics to calculate $\Delta \boldsymbol{u}_t$, but the value of $\Delta \boldsymbol{u}_t$ does not necessarily need to have a physical interpretation. 
The extended domain serves as a pure mathematical tool to counteract the corrosion of domain and make the output $\tilde{\boldsymbol{u}}_{t+\Delta t}$ satisfy the BC.
Therefore, the value of $\Delta \boldsymbol{u}_t$ can be freely searched from the range of real numbers.
In the view of optimization theory, imposing boundary condition by assigning values on the extended domain corresponds to an optimization problem: 
with the pre-trained LNO $\mathcal{G}_\theta$, the present flow field $\boldsymbol{u}_t$, boundary condition $\boldsymbol{u}=\boldsymbol{u}_\text{BC}$ on $\Gamma_\text{N}$ known, find a function $\Delta \boldsymbol{u}_t$ on $\Omega_\text{N}$ to minimize the object function
\begin{equation}
  \|\mathcal{G}_\theta(\boldsymbol{u}_t+\Delta \boldsymbol{u}_t)-\boldsymbol{u}_\text{BC}\|_{\Gamma_\text{N}}.
  \label{eq:3:boundary_loss}
\end{equation}
Here we propose a new strategy based on the backpropagation of neural networks to calculate $\Delta \boldsymbol{u}_t$.
The problem described in Eq. (\ref{eq:3:boundary_loss}) is built upon the pre-trained LNO, allowing the backpropagation to utilize of the learned weights $\theta$ of LNO and enhance optimization efficiency. 
The following gives the detailed steps.

At the beginning of optimization, set an initial value $\Delta \boldsymbol{u}_t^{(s=0)}$, where the superscript ‘$(s)$’ denotes the number of optimization iterations. 
An average filter with kernel size $k$ is applied to $\Delta \boldsymbol{u}_t^{(s)}$:
\begin{equation}
  \Delta \bar{\boldsymbol{u}}^{(s)}_t(x,y)=\frac{1}{k^2}\sum_{i,j=-\frac{k-1}{2}}^{\frac{k-1}{2}}\Delta \boldsymbol{u}^{(s)}_t(x+i\Delta x,y+j\Delta x).
  \label{eq:3:filter}
\end{equation}
This step reduces non-physical oscillation caused by the extended domain, which is discussed in detail in \ref{app:hyper_param}. 
In our code implementation, Eq. (\ref{eq:3:filter}) is conveniently realized through a convolutional layer with a kernel width of $k$ and constant weights of $\frac{1}{k^2}$.

Next, add $\boldsymbol{u}_t$ and $\Delta \bar{\boldsymbol{u}}^{(s)}_t$ as the input of the pre-trained LNO to obtain the predicted flow field $\tilde{\boldsymbol{u}}^{(s+1)}_{t+\Delta t}$:
\begin{equation}
  \tilde{\boldsymbol{u}}_{t+\Delta t}^{(s+1)}=\mathcal{G}_{\theta}(\boldsymbol{u}_t +h\Delta \bar{\boldsymbol{u}}^{(s)}_t).
  \label{eq:3:op_inference_s}
\end{equation}
The effect of $\Delta \bar{\boldsymbol{u}}^{(s)}_t$ is limited in $\Omega_\text{N}$ by multiplying $h$.

Then, extract the boundary value $\tilde{\boldsymbol{U}}^{(s+1)}_{t+\Delta t}$ from $\tilde{\boldsymbol{u}}^{(s+1)}_{t+\Delta t}$ and compute the discrepancy from the BC $\boldsymbol{u}_\text{BC}$ as the loss function $\mathcal{L}_\text{BC}$:
\begin{equation}
  \mathcal{L}=\frac{1}{N_\text{G}^{\text{Lagrange}}}\sum_{j=0}^{N_\text{G}^{\text{Lagrange}}}\left |\tilde{\boldsymbol{U}}_{t+\Delta t}^{(s+1)}(X_j)-\boldsymbol{u}_{\text{BC}}(X_j) \right |,
  \label{eq:3:op_loss_function}
\end{equation}
where the capital letters $\boldsymbol{U},\ \boldsymbol{X}$ emphasize they are values on Lagrangian points on the boundary. 
This notation aligns with that used in IBM mentioned in Section \ref{sec3:2:1}, making Eq. (\ref{eq:3:op_loss_function}) compatible with cases with both conforming and non-conforming meshes that $\tilde{\boldsymbol{U}}^{(s+1)}_{t+\Delta t}$ can be interpolated from $\tilde{\boldsymbol{u}}^{(s+1)}_{t+\Delta t}$ using Eq. (\ref{eq:3:IBM_begin}).

In practice, the discrepancy on the normal direction $\boldsymbol{n}$ and tangential direction $\boldsymbol{\tau}$ is counted separately as:
\begin{equation}
  \mathcal{L}_\text{BC}=\omega_n\mathcal{L}_n+\left(1-\omega_n\right)\mathcal{L}_\tau,
\end{equation}
with
\begin{equation}
  \mathcal{L}_n=\frac{1}{N_G^\text{Lagrange}}\sum_{j=1}^{N_G^\text{Lagrange}}\left|\left(\tilde{\boldsymbol{U}}^{(s+1)}_{t+\Delta t}(\boldsymbol{X}_j)-\boldsymbol{u}_\text{BC}(\boldsymbol{X}_j)\right)\cdot \boldsymbol{n}\right|,
\end{equation}
\begin{equation}
  \mathcal{L}_{\tau}=\frac{1}{N_G^\text{Lagrange}}\sum_{j=1}^{N_G^\text{Lagrange}}\left|\left(\tilde{\boldsymbol{U}}^{(s+1)}_{t+\Delta t}(\boldsymbol{X}_j)-\boldsymbol{u}_\text{BC}(\boldsymbol{X}_j)\right)\cdot \boldsymbol{\tau}\right|.
\end{equation}
$\omega_n$ is a constant to adjust the proportion of $\mathcal{L}_n$ and $\mathcal{L}_\tau$. 
We set $\omega_n=0.8$ to emphasize the importance of normal direction in the imposition of solid wall boundary condition. 
Usually, the tangential velocity quickly recovers to the mainstream value just a short distance from the boundary, whereas the normal velocity remains a small value. 
This implies that if a boundary error occurs in the normal direction, the relative error is much larger than that in the tangential direction, which has a much more detrimental effect on the predicted flow field. 
Setting $\omega_n>0.5$ helps to reduce the effect.

Finally, backpropagate to compute the gradient of $\mathcal{L}_\text{BC}$ and do one step optimization to update the  value on the extended domain, denoted as $\Delta\boldsymbol{u}^{(s+1)}_{t}$. 
Repeat this process until $\mathcal{L}_\text{BC}$ falls below a given threshold or the number of iterations $s$ reaches the setting upper limit. 
This is the end of one-step time-marching, and the predicted flow field at $t+\Delta t$ is $\tilde{\boldsymbol{u}}_{t+\Delta t}=\tilde{\boldsymbol{u}}^{(s+1)}_{t+\Delta t}$. 

The above process takes $\boldsymbol{u}_t$ on $\Omega$ as input, satisfying the IC; 
predicts through the pre-trained LNO, satisfying the governing equation; 
minimizes $\mathcal{L}_\text{BC}$ by optimization, satisfying the BC;
the input and output are both on the computational domain $\Omega$.
All four elements that decide the solution are clear, therefore, the output $\tilde{\boldsymbol{u}}_{t+\Delta t}$ should be the reliable prediction for the flow field at $t+\Delta t$. 
It is important to note that the available range of the strategy is not only limited to the solid wall BC discussed here. 
The strategy can be applied to various BCs, provided they can be formulated within the loss function Eq. (\ref{eq:3:op_loss_function}).

\subsubsection{Remark on different strategies}
\label{sec3:2:4}
The primary difference between the first two strategies (direct imposition and pressure symmetry) and the optimization-based algorthm is that, the first two strategies impose BCs on the input of LNO, whereas the opimization-based strategies aims to make the output satisfy the BC.
This implies an inevitable problem for direct imposition and pressure symmetry: the governing equation and the BC are not satisfied at the same time. 
The flow field after extension and correction (the middle pattern in Fig. \ref{fig_vde_process}) satisfies the BC but maybe not the governing equation. 
The output flow field (the right pattern in Fig. \ref{fig_vde_process}) is predicted by the pre-trained LNO which learned the fluid dynamics from data, thus it satisfies the governing equations but does not necessarily satisfy the BC. 
That is to say, the imposition of boundary condition and time marching are implemented in an asynchronous way, which leads to prediction error. 
Therefore, the direct imposition and pressure symmetry are collectively named asynchronous strategies.
The optimization-based strategy is also called the synchronous strategy.

\begin{figure}[htbp]
  \centering
  \includegraphics[width=0.6\textwidth]{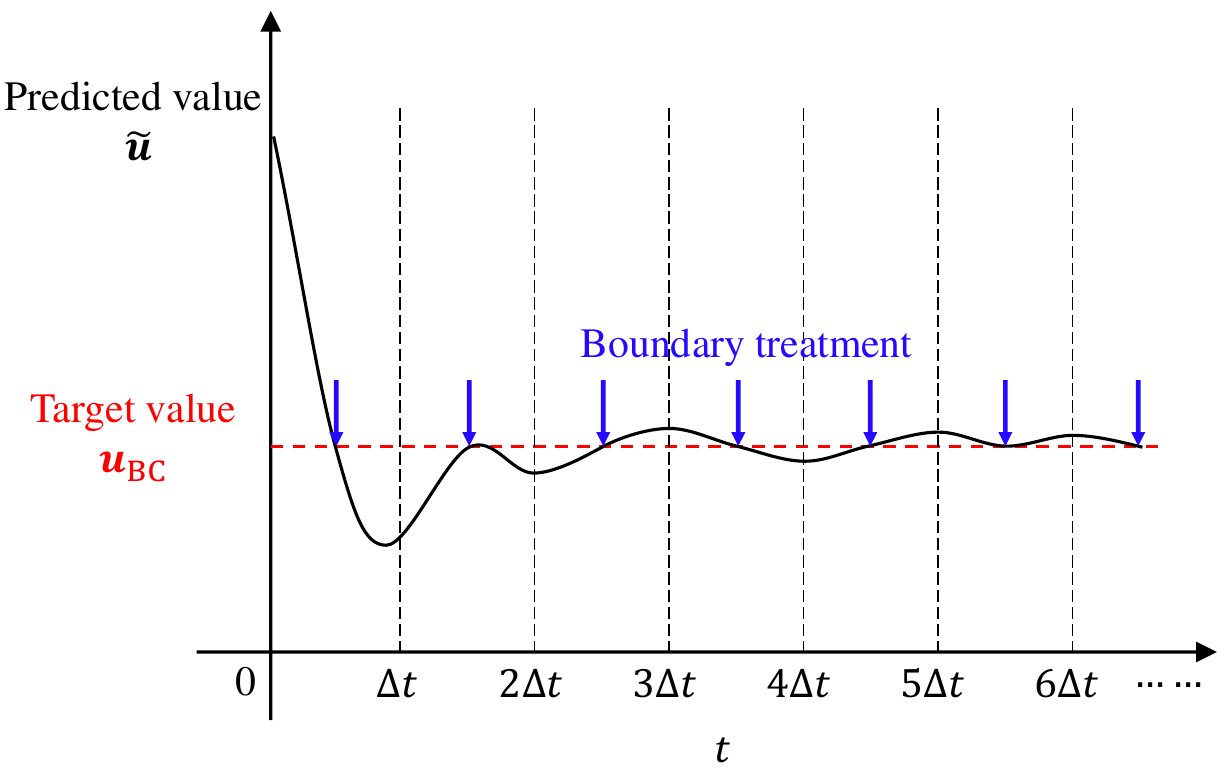}
  \caption{Schematic for the error caused by the asynchronous imposition of boundary conditions  in the time-marching prediction process}\label{fig_boundary_imposing_error}
\end{figure}

Fig. \ref{fig_boundary_imposing_error} illustrates the error caused by the asynchronous boundary treatment. 
The red dashed line represents the given value of BC $\boldsymbol{u}_\text{BC}$, and the black solid line represents the real value $\tilde{\boldsymbol{u}}$ at the boundary in LNO prediction. 
The blue arrows mark the moment of the boundary treatment, where $\tilde{\boldsymbol{u}}$ is forced to be the same as $\boldsymbol{u}_\text{BC}$. 
But in the subsequent time-marching process, $\tilde{\boldsymbol{u}}$ gradually deviates from $\boldsymbol{u}_\text{BC}$ until the next boundary treatment. 
The proximity of the two lines reflects the accuracy of imposing boundary condition. 
The error is neglectable when $\Delta t$ is small, which is the common case for traditional solvers.
However, it becomes considerable in LNO which featured for the large time interval.
The difference of asynchronous and synchronous strategies will be further illustrated by numerical examples in Section \ref{sec4}.

\subsection{Workflow}
\label{sec3:3}
\begin{figure}[htbp]
  \centering
  \includegraphics[width=0.85\textwidth]{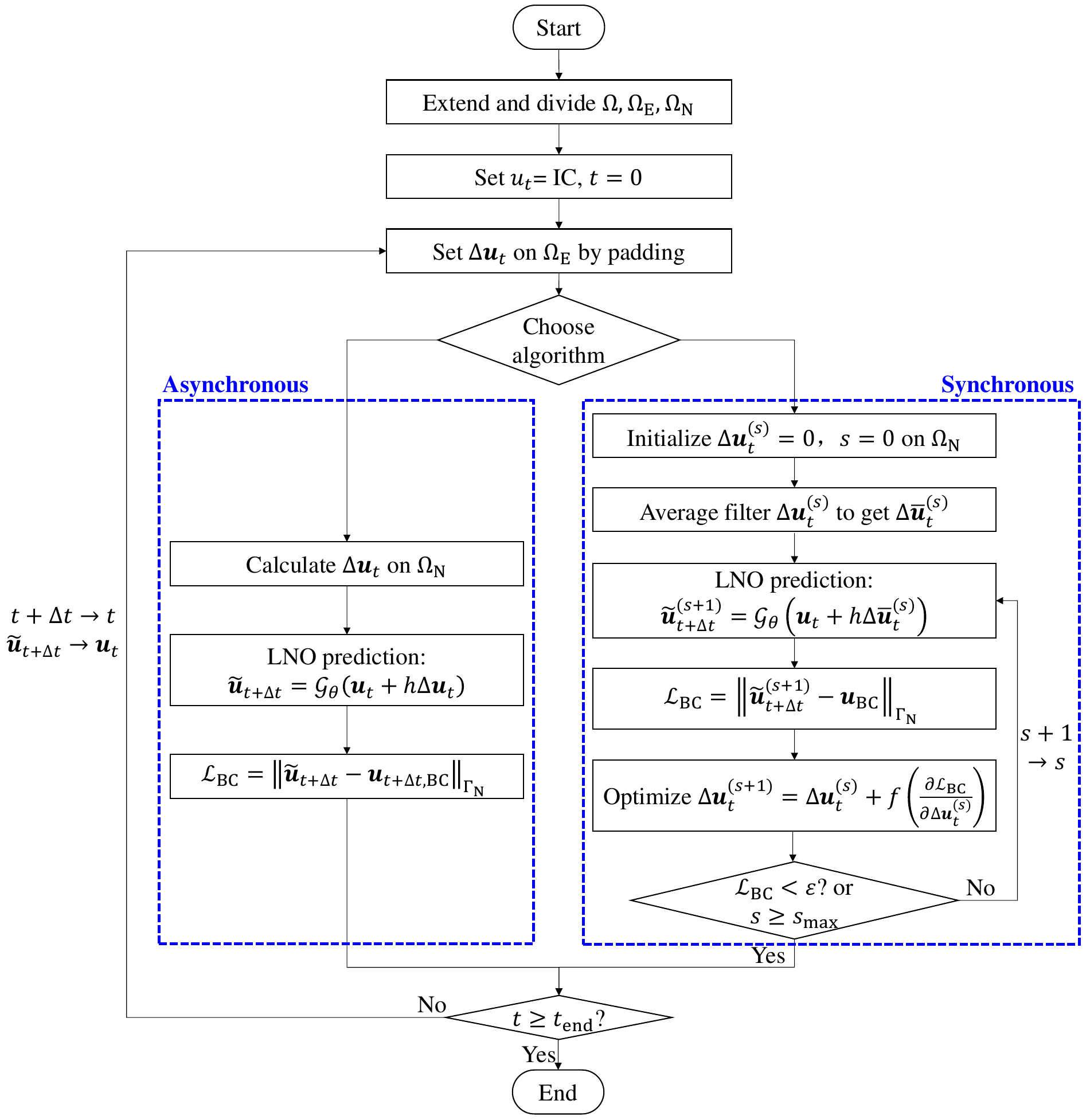}
  \caption{Flow chart for time-marching prediction of pre-trained LNO with boundary treatment. $f$ in the synchronous strategy depends on the chosen optimizer.}\label{fig_flow_chart}
\end{figure}

The workflow for predicting a flow using the pre-trained LNO and the boundary treatment is summarized in Fig. \ref{fig_flow_chart}. 
Before the calculation, the computational domain is extended and divided into three regions: 
$\Omega$,$\Omega_\text{E}$,$\Omega_\text{N}$, according to the type of boundary conditions as shown in Fig. \ref{fig_vde_process}. 
The extended width is set as the corrosion width of LNO to guarantee the domain remains unchanged after extension and corrosion. 
The time-marching prediction starts with the initial condition $\boldsymbol{u}_t$. 
To impose the extendable boundary conditions, fill $\Delta\boldsymbol{u}_t$ on $\Omega_\text{E}$ with padding operation. 
The specific type of padding is detailed in Fig. \ref{fig_padding} and Table \ref{tab_padding_option}. 

Then, to impose the non-extendable boundary conditions, one can choose from the asynchronous (direct imposition or pressure symmetry) and synchronous (optimization based on backpropagation) strategies. 
All the strategies are summarized in Table \ref{tab_algorithm_summary}.
When using the asynchronous one, calculate the velocity, the temperature, and  the density (if choose pressure symmetry) on $\Omega_\text{N}$. 
Put $\boldsymbol{u}_t$ and $\Delta\boldsymbol{u}_t$ together as the input of the pre-trained LNO to get the output $\tilde{\boldsymbol{u}}_{t+\Delta t}$. 
Then, calculate the difference between $\tilde{\boldsymbol{u}}_{t+\Delta t}$ and the boundary condition $\boldsymbol{u}_\text{BC}$ as $\mathcal{L}_\text{BC}$ to reflect the accuracy of boundary treatment. 
When using the synchronous strategy, first set the initial flow field on $\Omega_\text{N}$ as $\Delta \boldsymbol{u}^{(s)}_t=\boldsymbol{u}_\text{BC},s=0$.
Then use the averaged filter on $\Delta \boldsymbol{u}^{(s)}_t$ to get $\Delta \bar{\boldsymbol{u}}^{(s)}_t$.
Concatenate $\Delta \boldsymbol{u}^{(s)}_t$ and $\boldsymbol{u}_t$ and send to the pre-trained LNO to output $\tilde{\boldsymbol{u}
 }^{(s+1)}_{t+\Delta t}$. 
Calculate the difference between $\tilde{\boldsymbol{u}}^{(s+1)}_{t+\Delta t}$ and the boundary condition as the loss function $\mathcal{L}_\text{BC}$. 
Backpropagate and update the flow field on $\Omega_\text{N}$. 
Denote the updated field as $\Delta \boldsymbol{u}^{(s+1)}_t$. 
This completes a round of optimization. 
Check if the criterion $\mathcal{L}_\text{BC}<\varepsilon$ is satisfied or the round reaches the set maximum $s_\text{max}$. 
If not, start the next round of optimization. 
If yes, end the iteration and use $\tilde{\boldsymbol{u}}^{(s+1)}_{t+\Delta t}$ as the prediction for flow field at $t+\Delta t$, i.e., $\tilde{\boldsymbol{u}}_{t+\Delta t}$.

After a time-marching step, check if the moment $t$ reaches the set end time $t_\text{end}$. 
If not, use $\tilde{\boldsymbol{u}}_{t+\Delta t}$ as the initial condition of the next step, until $t\geq t_\text{end}$.


\begin{table}[htbp]
  \centering
  \caption{Summary of boundary treatment strategies}
  \label{tab_algorithm_summary}
  \begin{tabular}{cccc}
  \toprule
  Type                          & Strategy                                  & Formula for $T,\boldsymbol{v}$  & Formula for $\rho$ \\
  \midrule
  \multirow{3}{*}{Asynchronous} & \makecell[c]{Direct imposition on boundary curve\\(interior BC only)}                 &   $\Delta\boldsymbol{u}_t=\boldsymbol{f}_\text{N}\Delta t$ & / \\
  \cline{2-4}
                                & Direct imposition                  &  $\Delta\boldsymbol{u}_t=h\boldsymbol{f}_\text{N}\Delta t+h\left(\boldsymbol{u}_\text{BC}-\boldsymbol{u}_t\right)$  &  $\Delta\rho_t=1$ \\
                                \cline{2-4}
                                & Pressure symmetry                  &  $\Delta\boldsymbol{u}_t=h\boldsymbol{f}_\text{N}\Delta t+h\left(\boldsymbol{u}_\text{BC}-\boldsymbol{u}_t\right)$  & $\Delta \rho_A=h\frac{\rho_{A^\prime}T_{A^\prime}}{\Delta T_A}$       \\
  \midrule
  Synchronous                   & Optimization based on backpropagation &   \multicolumn{2}{c}{By optimization}         \\
  \bottomrule
  \end{tabular}
\end{table}

\section{Numerical examples}
\label{sec4}
This section presents three numerical examples that utilize the pre-trained LNO and boundary treatment strategies to compare the effectiveness of these strategies in predicting flow fields. 
The examples include an internal flow, the plane Poiseuille flow, and two external flows: 
flow around a circular cylinder and flow around a truck.

Before diving into these examples, we must first identify the indicators for whether the imposition of boundary conditions is good in the prediction of pre-trained LNO. 
Undoubtedly, the first indicator is the boundary error, represented by the loss function $\mathcal{L}_\text{BC}$ and the error distribution on the boundary. 
A smaller value of $\mathcal{L}_\text{BC}$ indicates a more accurate imposing of the boundary condition. 
Secondly, the overall prediction error on the computational domain can be taken as an indicator. 
The boundary condition largely affects the flow field, so a good imposition of the boundary condition relates to a good prediction. 
The last indicator we adopt is the streamline, as it allows for easy observation of whether the impermeable conditions are accurately imposed on solid walls. 
The first two indicators are in favor of quantitative evaluation, but they may be interfered by the inherent error of the pre-trained LNO, which brings uncertainty to the evaluation.
A large  $\mathcal{L}_\text{BC}$ might correspond to small overall prediction error, and conversely, a smaller value of $\mathcal{L}_\text{BC}$ does not necessarily guarantee a better prediction.
Readers refer to \ref{app:hyper_param} where an example of small $\mathcal{L}_\text{BC}$ leading to bad prediction is provided. 
Here comes the need for a non-quantitative indicator, the streamline, which is presented in visual form.
Its judgement for good predictions is straightforward: 
one simply needs to check if the streamline penetrates the solid wall. 
Overall, the three indicators should be considered comprehensively to assess the prediction.

\subsection{Plane Poiseuille flow}
\label{sec4:1}
\begin{figure}[htbp]
  \centering
  \includegraphics[width=0.75\textwidth]{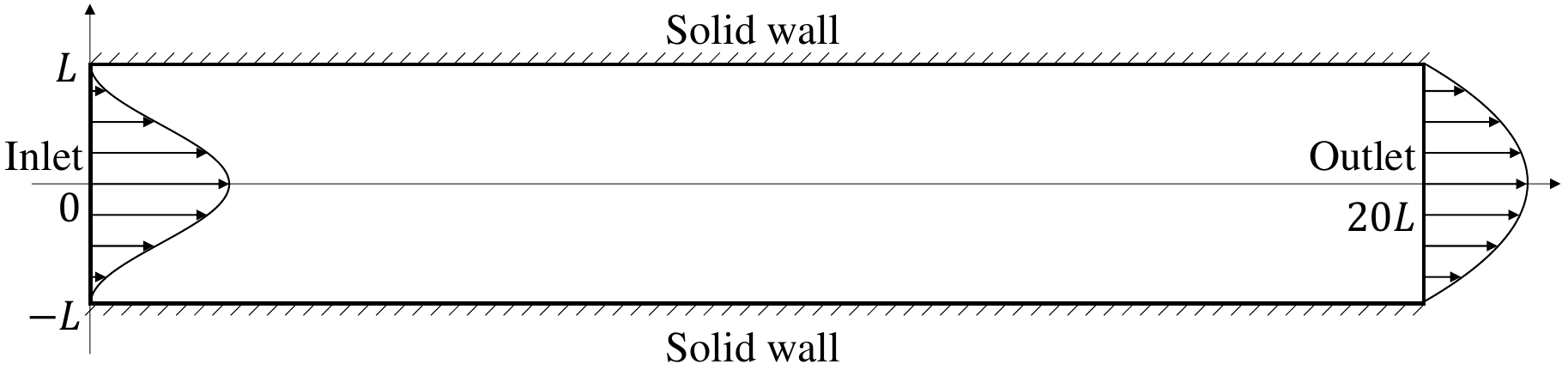}
  \caption{Schematic of the plane Poiseuille flow}\label{fig_poiseuille_schematic}
\end{figure}

The first numerical example is the plane Poiseuille flow. 
As shown in Fig. \ref{fig_poiseuille_schematic}, the fluid flows from the left of a straight tube to the right, and the velocity at the inlet is:
\begin{equation}
  v_x=\cos^2{\left(\frac{\pi y}{2L}\right),\ v_y=0}.
  \label{eq:4:poiseuille_inlet}
\end{equation}

After the flow is fully developed, there exists the analytical solution of the velocity, which is a parabolic function with respect to $y$ \cite{Poinsot1992}:
\begin{equation}
  v_x=\frac{3}{4L^2}\left(L^2-y^2\right).
  \label{eq:4:poiseuille_developed}
\end{equation}
Set isothermal velocity boundary conditions on all sides of the computational domain except for the outlet, wherein the upper and lower sides are static solid wall with $v_x=v_y=0,T=1$, the inlet velocity is given by Eq. (\ref{eq:4:poiseuille_inlet}) and $T=1,\rho=1$.
As for the outlet boundary, commonly the outflow BC for $\rho,T,\boldsymbol{v}$ is adopted, but here we change the BC for $\boldsymbol{v}$ to the known distribution Eq. (\ref{eq:4:poiseuille_developed}) to focus on examining the accuracy of the boundary treatment strategies.
Set $L=1$, the computational domain is $\left[0,20\right]\times[-1,1]$. 
The grid size of the pre-trained LNO is $\Delta x=1/64$, so the total number of Eulerian points is $N_G^\text{Euler}=1281\times129=165249$. 
As the four boundaries are all straight, the Lagrangian points are selected as the Eulerian points on the boundary, so $N_G^\text{Lagrange}=2816$. 
The initial condition assumes a uniform distribution of density and temperature, says $\rho=1$ and $T=1$. 
The velocity is set as the inlet distribution on the entire domain.

All the boundaries are non-extendable exterior ones, so we will compare three boundary treatment strategies in this example: 
the first is direct imposition, where $\boldsymbol{v}$, $T$ are set as the boundary values and $\rho$ is set as 1 on the extended domain; 
the second is pressure symmetry, which treats $\boldsymbol{v}$, $T$ the same as direct imposition and uses pressure symmetry to calculate $\rho$; the third is optimization-based strategy, which uses optimization based on backpropagation to determine the value on the extended domain. 

\begin{figure}[htbp]
  \centering
  \includegraphics[width=0.6\textwidth]{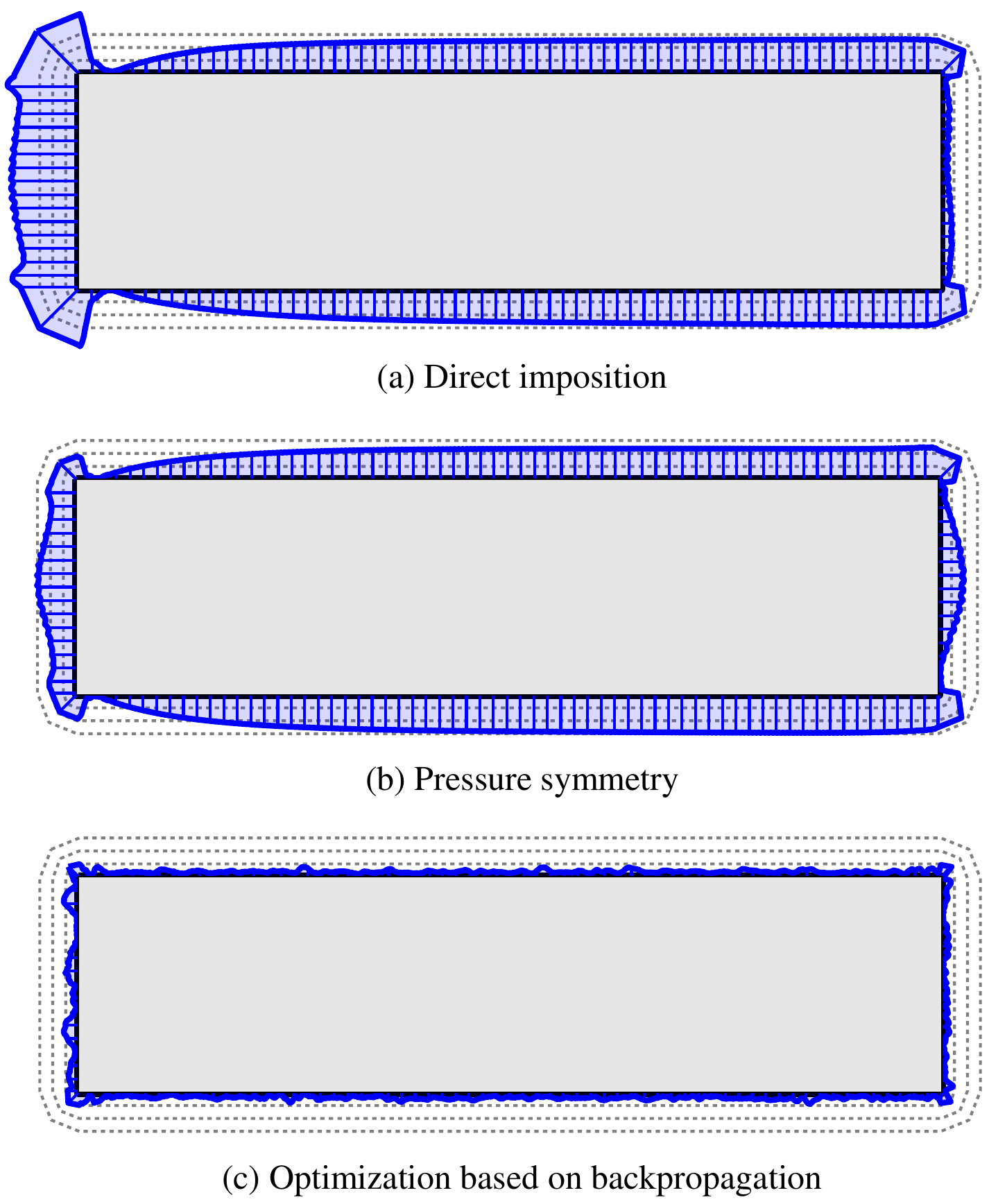}
  \caption{Plane Poiseuille flow: distribution for the magnitude of velocity error on the boundary of the tube. The $x$ axis is compressed by a factor of $2.5$ for better exhibition. The three gray dotted squares denote error of magnitude $0.01$, $0.02$, and $0.03$, respectively.}\label{fig_poiseuille_error_dist}
\end{figure}

\begin{figure}[htbp]
  \centering
  \includegraphics[width=0.5\textwidth]{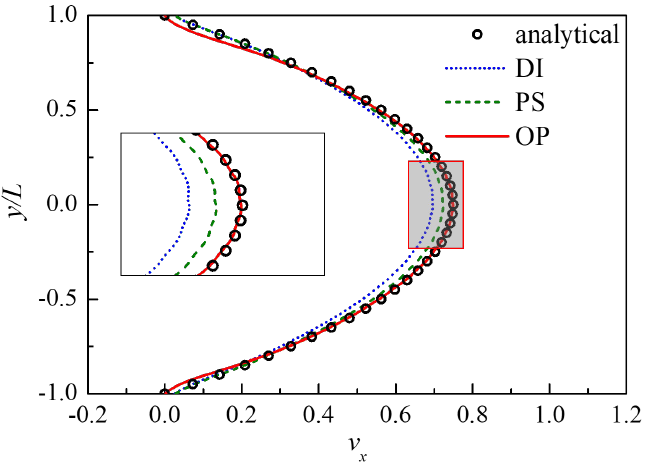}
  \caption{Plane Poiseuille flow: velocity profile at $x=\frac{40}{3}L$ and comparison with the analytical solution.DI, PS, OP stand for direct imposition, pressure symmetry, and optimization-based strategy, respectively.}\label{fig_poiseuille_profile}
\end{figure}

Fig. \ref{fig_poiseuille_error_dist} and Table \ref{tab_error_list} exhibit the distribution and averaged value of error on the tube boundary after the flow is fully developed. 
The results show that the imposition of boundary conditions by optimization-based strategy is the most accurate, followed by pressure symmetry, and direct imposition is the worst. 
The imposition of boundary conditions affects not only the flow field near the boundary, but also the interior flow away from the boundary. 
Fig. \ref{fig_poiseuille_profile} exhibits the velocity profiles at the middle of the tube after the flow is fully developed and compares them with the analytical solution. 
The velocity profile by optimization-based strategy aligns well with the analytical solution, while the profiles of the other two strategies are lower in the middle.

\begin{table}[htbp]
  \centering
  \caption{The averaged error with different boundary treatment strategies}
  \label{tab_error_list}
  \begin{tabular}{cccc}
  \toprule
  \multirow{2}{*}{Problem}                                 & \multirow{2}{*}{Algorithm} & \multicolumn{2}{c}{Boundary error} \\
                                                    &  &  $\mathcal{L}_{\tau}$   &  $\mathcal{L}_{n}$     \\
  \midrule
  \multirow{3}{*}{Plane Poiseuille flow}                   & Direct imposition       & 0.0212 & 0.0035 \\
                                                           & Pressure symmetry        & 0.0215 & 0.0019 \\
                                                           & Optimization-based strategy        & 0.0024 & 0.0031 \\
                                                           \midrule
  \multirow{4}{*}{\makecell[c]{Flow around a circular cylinder\\($t=1.4$)}} & Direct imposition on boundary curve      & 0.3847 & 0.4388 \\
                                                           & Direct imposition       & 0.1597 & 0.0861 \\
                                                           & Pressure symmetry        & 0.1543 & 0.0843 \\
                                                           & Optimization-based strategy        & 0.0708 & 0.0510 \\
                                                           \midrule
  \multirow{4}{*}{\makecell[c]{Flow around a truck\\($t=28$)}}              & Direct imposition on boundary curve      & 0.2429 & 0.1535 \\
                                                           & Direct imposition       & 0.0749 & 0.0225 \\
                                                           & Pressure symmetry        & 0.0739 & 0.0151 \\
                                                           & Optimization-based strategy        & 0.0679 & 0.0097 \\
  \bottomrule
  \end{tabular}
  \end{table}

\subsection{Flow around a circular cylinder}
\label{sec4:2}
\begin{figure}[htbp]
  \centering
  \includegraphics[width=\textwidth]{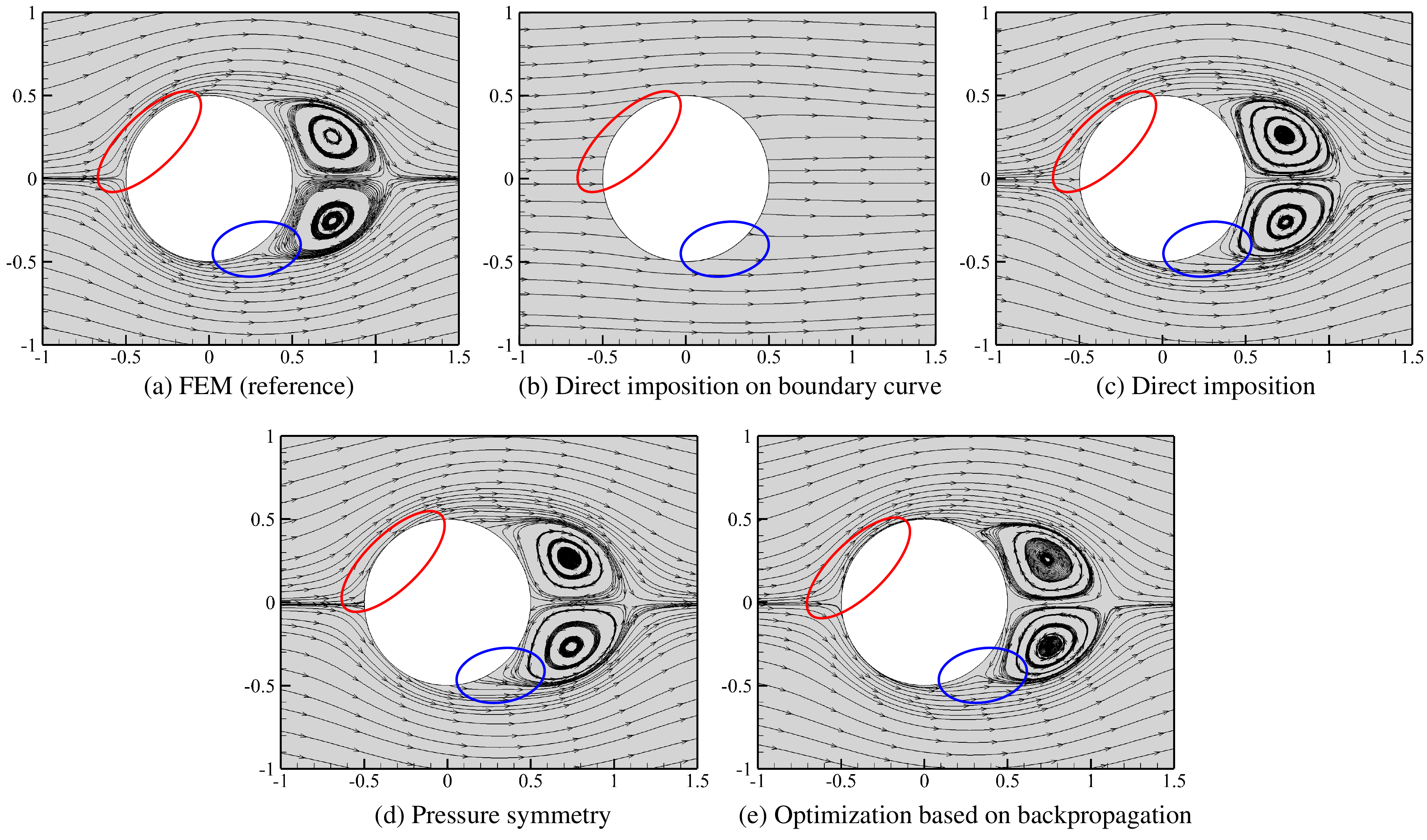}
  \caption{Flow around a circular cylinder: streamlines around the cylinder at $t=1.4$. The two circles mark the main differences of different strategies.}\label{fig_CC_streamline}
\end{figure}

A circular cylinder with diameter $D=1$ is placed in 2-D infinite plane, and the uniform flow $v_x=1$ comes from the left. 
Here we mainly focus on the imposition of the boundary condition on the cylinder wall, which is non-extendable interior boundary condition. 
As discussed in Section \ref{sec3:2:1}, when using direct imposition, one can choose from only correcting the annular domain within $\Delta x$ from the boundary (direct imposition on boundary curve) or, on this basis, setting the interior region to the boundary value (direct imposition). 
The performance of the two strategies will be testified and compared in this case. 
Hence, totally four strategies (direct imposition on boundary curve, direct imposition, pressure symmetry, and optimization-based strategy) are discussed below.
Fig. \ref{fig_CC_streamline} demonstrates the streamline near the circular cylinder at $t=1.4$ where a pair of vortices are generated. 
The streamline by FEM is also exhibited in Fig. \ref{fig_CC_streamline} (a) as reference. 

The first concern is the difference between the two direct imposition strategies. 
Clearly, the result by direct imposition on boundary curve largely deviates from the reference. 
The streamlines pass through the solid wall of the cylinder freely with only a slight bending of streamlines occurs downstream of the cylinder. 
On the contrary, direct imposition brings out roughly correct results. 
The only distinction between the implementation of direct imposition on boundary curve and direct imposition lies in whether the enclosed region of the cylinder is set as the boundary value. 
In traditional solvers with small time intervals, they make little difference, however, in LNO prediction with large time interval, the difference of the two strategies becomes remarkable. 
The observation indicates the large time interval of LNO requires more for the boundary treatment strategy that the value of every node on the extended domain must be carefully assigned to ensure a good prediction.

Next, we compare the asynchronous and synchronous strategies. 
Two main differences are marked with circles in Fig. \ref{fig_CC_streamline}. 
The first is located on the left of the cylinder (marked as the red circle), which directly faces the coming stream.
It requires the velocity to decrease rapidly from $1$ to $0$ and challenges the boundary treatment. 
In asynchronous strategies (direct imposition and pressure symmetry), we observe streamline penetration, whereas in the synchronous strategy, the streamline successfully bypasses the cylinder. 
The second is near the right upper and lower sides of the cylinder, where the streamline tends to develop along the cylinder wall, but soon bends outward due to the flow separation. The small bend is only captured by optimization-based strategy.

\begin{figure}[htbp]
  \centering
  \includegraphics[width=\textwidth]{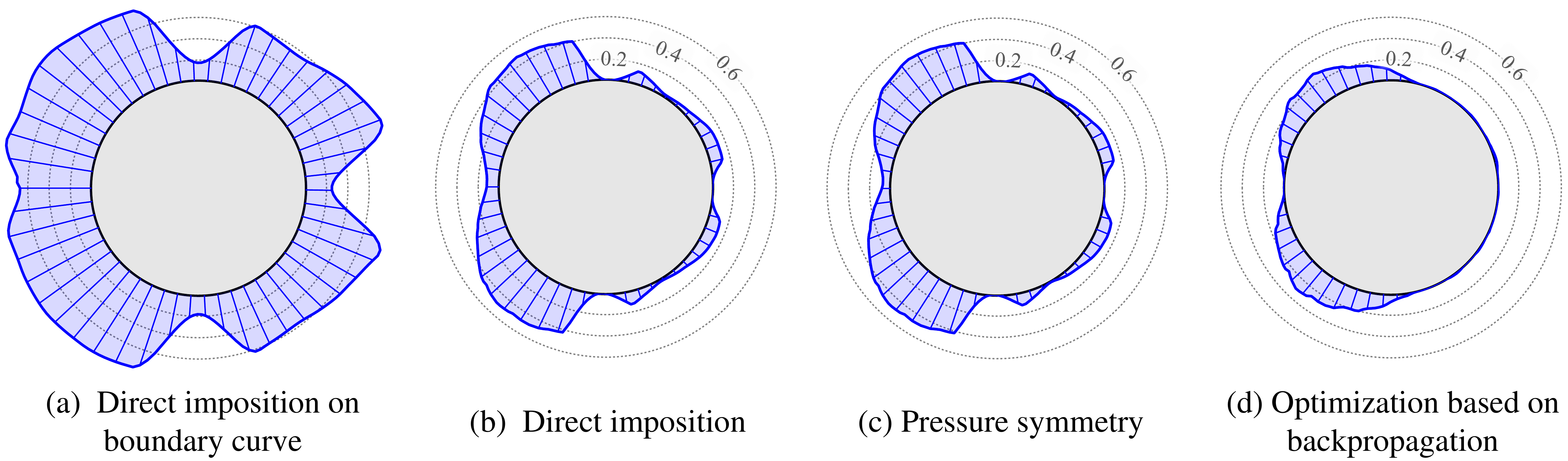}
  \caption{Flow around a circular cylinder: distribution for the magnitude of velocity error on the cylinder wall. The three gray dotted circles denote error of magnitude $0.2$, $0.4$, and $0.6$, respectively.}\label{fig_CC_error_dist}
\end{figure}

Fig. \ref{fig_CC_error_dist} exhibits the error distribution on the cylinder wall. 
The length of each straight blue line represents the magnitude of velocity error $\sqrt{v_x^2+v_y^2}$ at that specific point. 
The error on the left half of the cylinder is large, because it is the upwind side where the velocity decreases rapidly.
The error distribution corresponds well with the earlier observation on streamlines: 
the error of direct imposition on boundary curve is the largest, followed by direct imposition and pressure symmetry, while optimization-based strategy has the smallest error. 
direct imposition and pressure symmetry do not show much difference in the streamline and error distribution in this case, but the quantitative value of the velocity error listed in Table \ref{tab_error_list} indicates there is a slight improvement in both tangential and normal errors from direct imposition to pressure symmetry.

\begin{figure}[htbp]
  \centering
  \includegraphics[width=\textwidth]{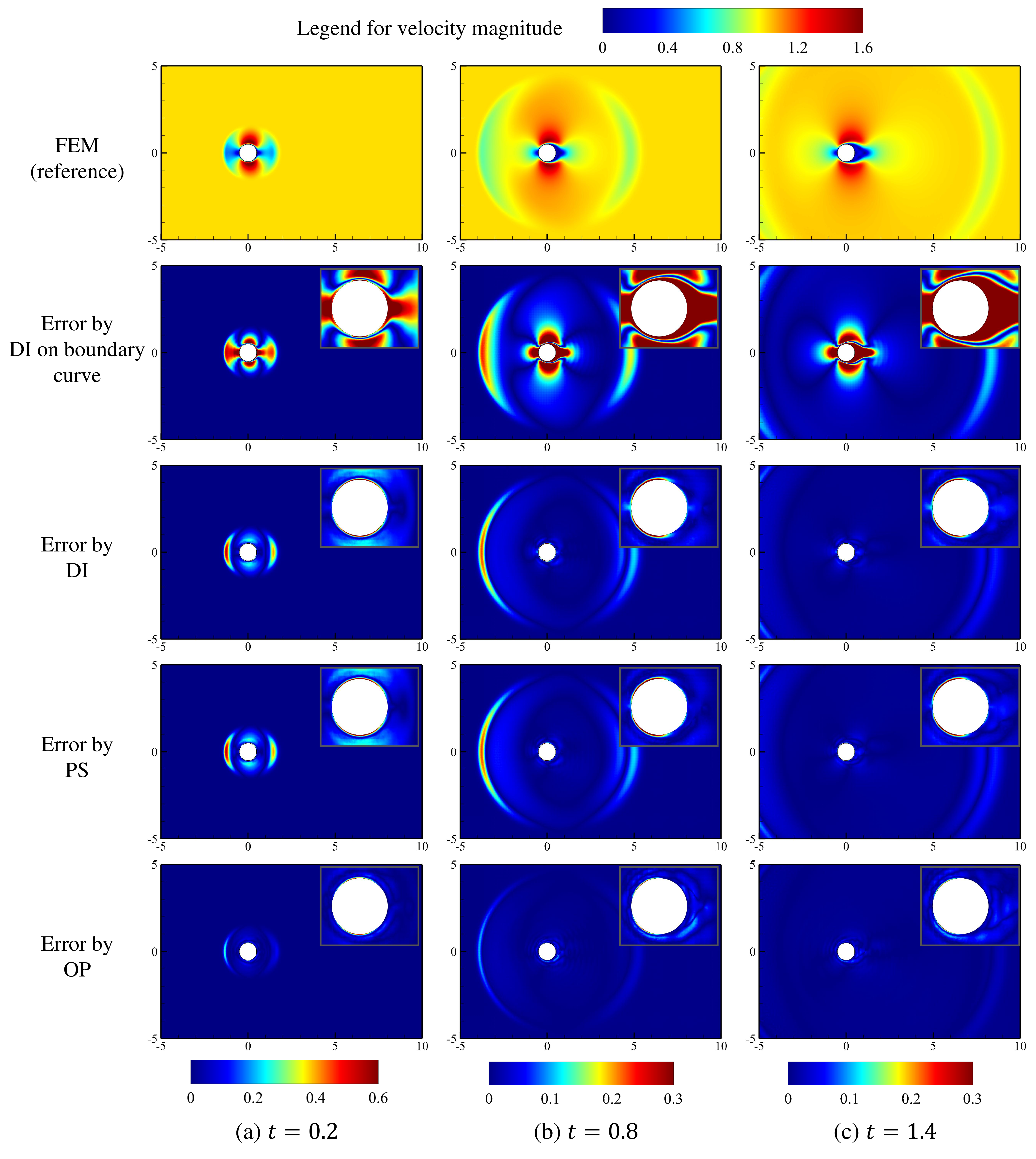}
  \caption{Flow around a circular cylinder: velocity magnitude and its absolute error by different boundary treatment strategies. The first row is velocity magnitude calculated by FEM, here are used as the reference for compute the absolute error. Other rows are the absolute error by different boundary treatment strategies. DI, PS, OP stand for direct imposition, pressure symmetry, and optimization-based strategy, respectively.}\label{fig_CC_error_contour}
\end{figure}

The error on the boundary spreads and whereafter becomes error on the computational domain, as shown in the contours in Fig. \ref{fig_CC_error_contour}. 
Algorithms with larger boundary error correspond to the contours displaying greater overall error. 
The issue is pronounced on the capture of the initial wave that generated by the cylinder wall at the starting moment and propagates at the sound speed. 
The accuracy of capturing this wave is highly influenced by how accurately the boundary condition is imposed. 
Consequently, larger errors are observed in the predicted flow fields of direct imposition on boundary curve, direct imposition, and pressure symmetry strategies.

\subsection{Flow around a vehicle}
\label{sec4:3}
The last case is the flow around a truck in the tunnel, which was presented in our previous paper \cite{Ye2023b}. 
Here we place emphasis on the boundary treatment for the vehicle wall.

\begin{figure}[htbp]
  \centering
  \includegraphics[width=0.8\textwidth]{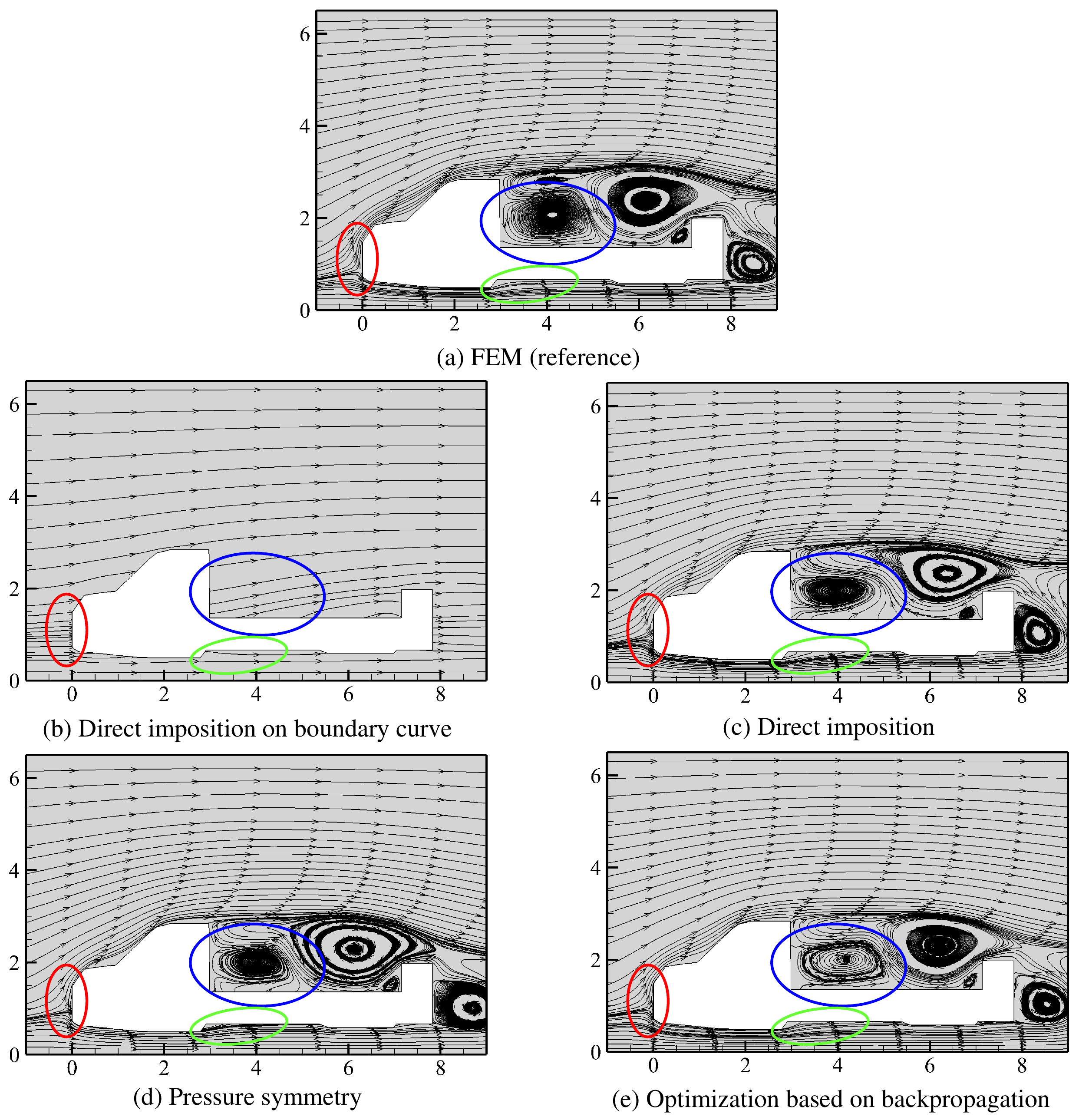}
  \caption{Flow around a vehicle: streamlines around the truck at $t=28$. The three circles mark the main differences of different strategies.}\label{fig_truck_streamline}
\end{figure}

Fig. \ref{fig_truck_streamline} exhibits the streamline around the vehicle at $t=28$. 
Similar to Section \ref{sec4:2}, the predicted flow field by direct imposition on boundary curve fully permeates into the truck and is far from the reference solution. 
The other three strategies succeed in capturing the vortices while their main differences are found in three places as indicated with circles in Fig. \ref{fig_truck_streamline}. 
The first is the leading edge of the truck (marked with a red circle), where the fluid should move around the solid wall. 
In the results by direct imposition and pressure symmetry, several streamlines dive into the boundary, showing the inaccuracy in imposition of the boundary condition, while the result by optimization-based strategy describes the streamlines correctly. 
The second difference occurs in the lower-left corner of the container (marked with a blue circle) where a counterclockwise vortex exists. 
For direct imposition and pressure symmetry, multiple streamlines are non-physically generated from the truck wall, whereas the optimization-based strategy largely alleviate this issue. 
The last difference is at the bottom of the truck (marked with a green circle) where the fluid should flow along the wall. 
Some streamlines are drawn into the truck in the figure of direct imposition. 
In contrast, pressure symmetry and optimization-based strategy both provide good predictions. 
This case clearly reveals the improvement from direct imposition to pressure symmetry, which is adept at handling the solid wall along the mainstream direction.

\begin{figure}[htbp]
  \centering
  \includegraphics[width=\textwidth]{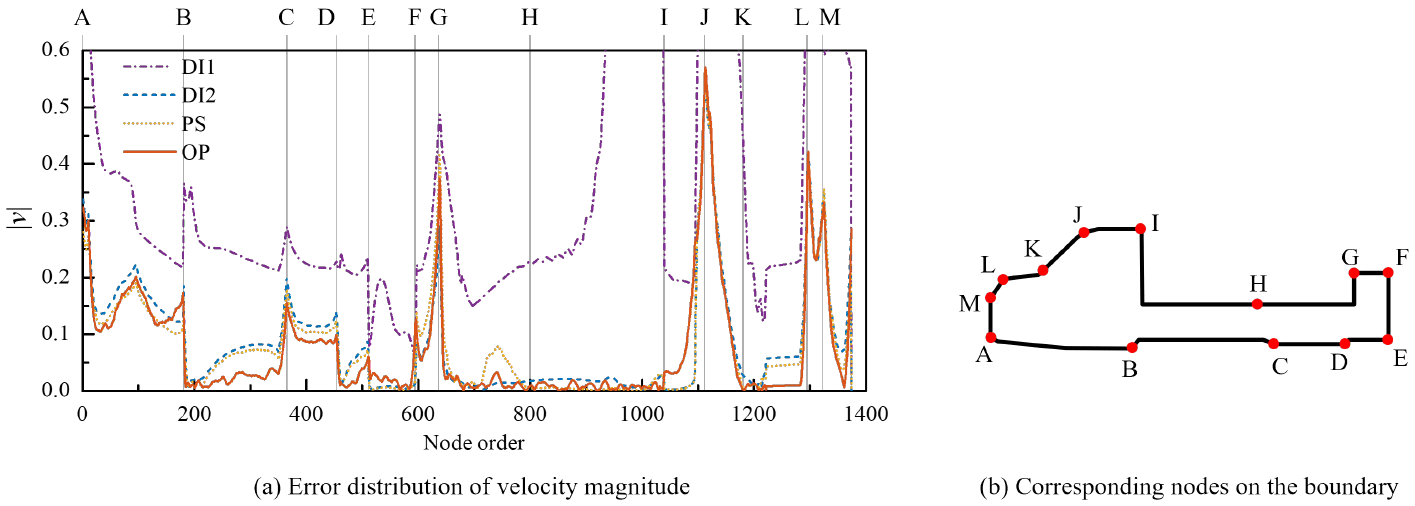}
  \caption{Flow around a vehicle: distribution for the magnitude of velocity error on the solid wall.}\label{fig_truck_error_dist}
\end{figure}

\begin{figure}[htbp]
  \centering
  \includegraphics[width=0.8\textwidth]{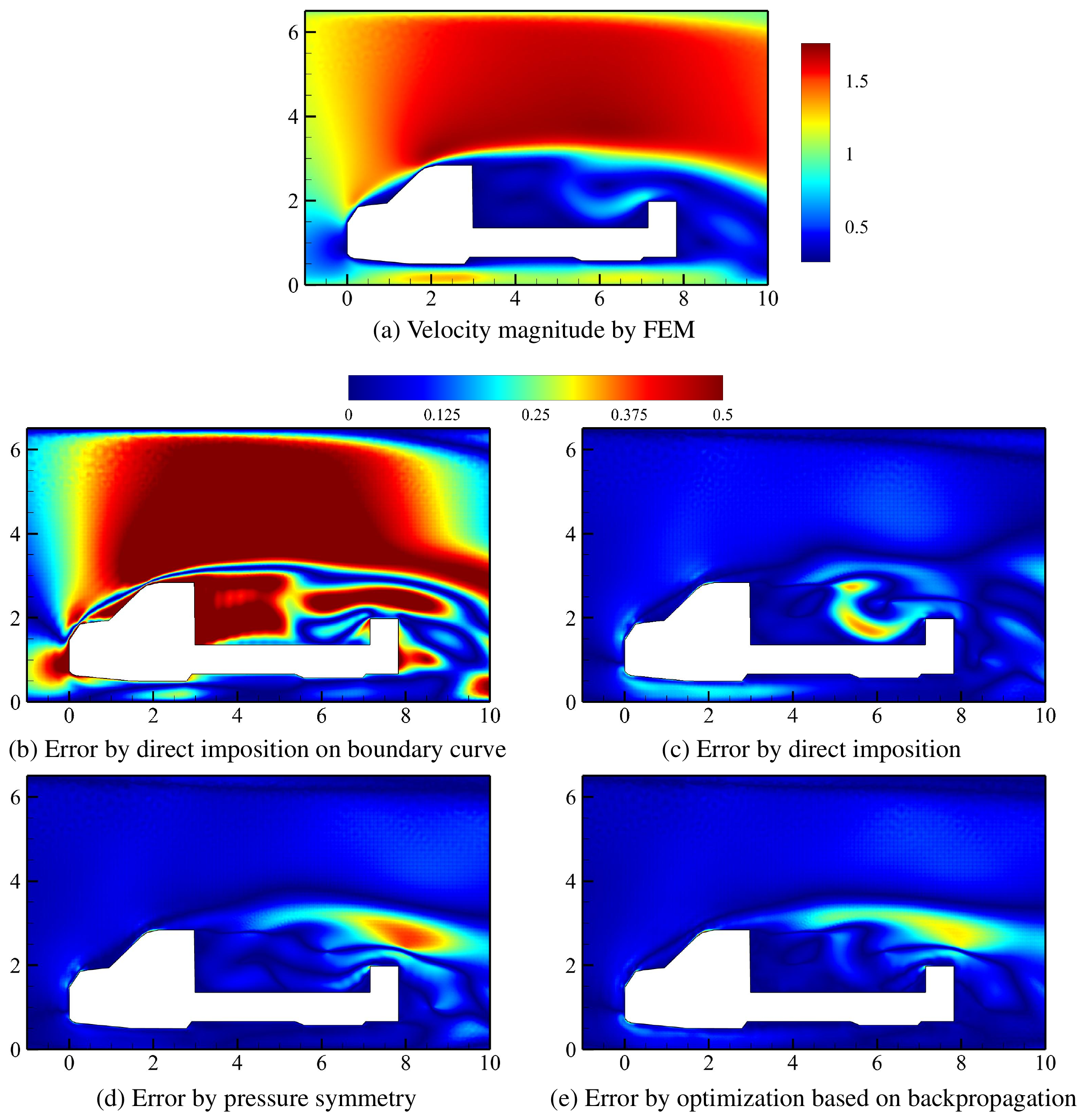}
  \caption{Flow around a truck: velocity magnitude and its absolute error by different boundary treatment strategies.}\label{fig_truck_error_contour}
\end{figure}

Fig. \ref{fig_truck_error_dist} provides the error distribution at the truck wall, then Fig. \ref{fig_truck_error_contour} exhibits the error distribution on the computational domain. 
Large errors mainly occur in the upwind turning points, such as A, G, J, L, M in Fig. \ref{fig_truck_error_dist}(b). 
Among them, point J on the top of the truck is with the largest error, because this point is near the zone with the highest velocity (refer to Fig. \ref{fig_truck_error_contour}(a)). 
It is consistent with the results in Section \ref{sec4:2} that the largest error occurs on the lower and upper left in the upwind side. 
Reducing the boundary error at these upwind points is difficult, especially in this case with sharp corners, however, the predicted flow fields around the corners of different strategies vary.
In Fig. \ref{fig_truck_error_contour}, the error around point A, L, G decreases in the order of direct imposition on boundary curve, direct imposition, pressure symmetry and optimization-based strategy.
In addition, the error at other points differs evidently in Fig. \ref{fig_truck_error_dist}, which also reflects in the averaged boundary error in Table \ref{tab_error_list}.

\vspace{5mm}
Based on the observations of these numerical examples, we could summarize the feature of the four boundary treatment strategies when collaborating with the pre-trained LNO. 
direct imposition on boundary curve is not available for use due to the mismatch with the large time interval of the pre-trained LNO. 
It makes the boundaries ignored in the prediction and leads to flow field like freestream. 
Direct imposition could provide roughly correct flow patterns, such as the number and positions of vortices. 
However, it exhibits a significant boundary error and the fluid non-physically penetrates the solid wall. 
Pressure symmetry is an improvement over direct imposition,  the density value is assigned by the pressure symmetry formulation. 
It reduces the boundary error and performs particularly well in the case with solid walls along the mainstream. 
The limitation of pressure symmetry is it only applies to the solid wall BC and can not handle other types of BCs. 
Finally, optimization-based strategy is the most accurate among all the strategies. 
It largely reduces the boundary error and ensures the streamlines pass through the solid wall physically. 
Moreover, optimization-based strategy is applicable in all types of boundary conditions that can be explicitly expressed as mathematic formulations, beyond just solid wall.

\section{Discussion and conclusion}
\label{sec5}
This paper discussed a method named virtual boundary extension (VDE) for imposing boundary conditions (BCs) in the prediction of fluid flows using pre-trained local neural operator (LNO). 
VDE leverages the `corrosion of domain' phenomenon in LNO and introduces the information of BCs by assigning proper values to the virtual domains extended from the boundaries. 
Within VDE framework, the main discussion is about how to determine the values to be assigned on the extended domain for different BCs.
BCs are categorized into extendable and non-extendable ones. 
For extendable ones such as the periodic and symmetric BCs, simple padding operations can impose the boundary conditions exactly. 
In contrast, non-extendable ones, represented by the solid wall BCs, require more complex treatment. 
Three strategies are proposed to calculate the assigned value on the extended domains including direct imposition (direct imposition on boundary curve and direct imposition on the whole extended domain), pressure symmetry, and optimization by backpropagation. 
These strategies are applied and compared through three numerical examples. 
The results show optimization-based strategy has the highest accuracy and widest applicable range, while pressure symmetry is easy to implement and works particularly well on straight solid walls aligned with the mainstream. 
This work provides LNO the resuablity of imposing various BCs, which serves as the last step to conceptually achieve the reusability of a single pre-trained LNO across problems with different problem-specific conditions.

The distinct nature of LNO compared to traditional solvers necessitates tailored boundary treatment strategies.
It is observed in the numerical examples that some strategies from traditional solvers (e.g., direct imposition on boundary curve) could hardly introduce the boundary effect to the predicted flow field when collaborating with LNO.
The primary cause is the large time interval $\Delta t$ of the pre-trained LNO leads to a wide affected domain near the boundary. 
Introducing the boundary effect to only the boundary curve or a small near-boundary range is insufficient to correctly impose BCs.
While these strategies may be effective again if we switch to another pre-trained LNO with smaller  $\Delta t$ to reduce the affected domain, reducing  $\Delta t$ would undermine one of LNO’s fundamental advantages: 
its high efficiency. 
Therefore, investigating boundary treatment strategies based on the characteristics of LNO is necessary.

An unexpected finding from numerical examples is that a smaller boundary error does not always lead to a better prediction. 
For example, in Fig. \ref{fig_truck_error_contour}, the velocity error on part of the computational domain produced by pressure symmetry or optimization-based strategy is even larger than that by direct imposition. 
The phenomenon could raise from the approximation error of the pre-trained LNO to the solution operator.
The final error on the computational domain is the superposition of LNO's approximation error and boundary error. 
Consequently, similar to the issue of overfitting in neural network training, paying too much attention to reducing the boundary error distorts the flow field near the boundary and causes an increase in the overall prediction error. 
Conversely, if the approximation of the pre-trained LNO is accurate enough, the implementation of VDE could be simpler and more straightforward.
By then, boundary error can be used as the only indicator of the optimization-based algorthm without causing mismatch between different errors.

The present work inspires several future directions of study on LNO. 
First, the strategy to calculate values on the extended domain could be further improved and explored within the VDE framework.
Second, there is a need to enhance the performance of the pre-trained LNO, focusing on aspects such as accuracy, extrapolation, and stability. 
Improving these foundational elements is crucial for accurate boundary condition imposition and reliable flow field prediction.
Then, LNO could be progressively moved from the present validation examples to more complex applications, allowing for the exploration of its full potential in practice.

\section*{Code availability}
The code is available at https://github.com/PPhub-hy/torch-virtual-domain-extension.


\nomenclature{$\boldsymbol{v}=\{v_x,v_y\}$}{Velocity with its two components }
\nomenclature{$\rho$}{Density}
\nomenclature{$T$}{Temperature}
\nomenclature{$p$}{Pressure}
\nomenclature{$E$}{Total energy}
\nomenclature{$\mu$}{Viscosity}
\nomenclature{$R$}{Gas constant}
\nomenclature{$C_\mathrm{v}$}{Heat capacity}
\nomenclature{$\kappa$}{Thermal conductivity}
\nomenclature{$Re$}{Reynolds number}
\nomenclature{$Ma$}{Mach number}
\nomenclature{$Pr$}{Prandtl number}
\nomenclature{$\gamma$}{Specific heat ratio}

\nomenclature{$\mathcal{G}_\mathrm{L}$}{Target local-related time-marching operator of transient PDEs}
\nomenclature{$\mathcal{G}_\theta$}{LNO for approximating $\mathcal{G}_\mathrm{L}$ with  the set $\theta$ of trainable weights}
\nomenclature{$\Omega$}{Computational domain}
\nomenclature{$\Gamma_\text{E}$}{Extendable boundary}
\nomenclature{$\Gamma_\text{N}$}{Non-extendable boundary}
\nomenclature{$\Omega_\text{E}$}{Extended domain from the extendable boundary $\Gamma_\text{E}$}
\nomenclature{$\Omega_\text{N}$}{Extended domain from the non-extendable boundary $\Gamma_\text{N}$}


\nomenclature{$\boldsymbol{u}_t$}{Physical fields at time $t$ on $\Omega$}
\nomenclature{$\tilde{\boldsymbol{u}}_{t+\Delta t}$}{The predicted physical fields at time $t+\Delta t$ by LNO on $\Omega$}
\nomenclature{$\Delta \boldsymbol{u}_t$}{Physical fields at time $t$ on the extended domain to impose boundary conditions}
\nomenclature{$h$}{Indicator of a point wether is on $\Omega_\text{N}$, $h=1$ denotes on $\Omega_\text{N}$ and $h=0$ denotes not}
\nomenclature{$\Delta \boldsymbol{u}_t^{(s)}$}{Physical fields in $s^{\text{th}}$ iteration of backpropagation optimization at time $t$ on the extended domain to impose boundary conditions}
\nomenclature{$s$}{Number of iterations of backpropagation optimization}

\nomenclature{$D_\text{out},D_\text{in}$}{Representational unit output/input domain in LNO definition}
\nomenclature{$\Omega_\text{out},\Omega_\text{in}$}{Complete output/input domain in LNO definition}
\nomenclature{$\mathcal{L}$}{Loss function for LNO training}
\nomenclature{$\mathcal{L}_\text{BC}$}{Loss function and the indicator for boundary error in imposition of boundary conditions}
\nomenclature{$\mathcal{L}_{n}$}{Loss function of boundary error in the normal direction}
\nomenclature{$\mathcal{L}_{\tau}$}{Loss function of boundary error in the tangential direction}
\nomenclature{$\omega_n$}{Coefficient to control the proportion of $\mathcal{L}_{n}$ and $\mathcal{L}_{\tau}$}
\nomenclature{$n$}{Block number}
\nomenclature{$N$}{Width of local spectral transform}
\nomenclature{$K$}{Number of repetitions for geometry decomposition}
\nomenclature{$M$}{Number of spectral modes adopted}

\nomenclature{$k$}{Kernel size of the average filter}
\nomenclature{$\lambda$}{Coefficient of weight decay in the loss function}
\nomenclature{$lr$}{Learning rate of the optimizer}

\nomenclature{$N_\text{G}^{\text{Euler}}$}{Total number of equidistant nodes in the computational domain}
\nomenclature{$N_\text{G}^{\text{Lagrange}}$}{Total number of  nodes on the boundary curve}
\nomenclature{$\boldsymbol{U}$}{Physical fields on the boundary curve}
\printnomenclature

\appendix
\section{Architecture of local neural operator}
\label{app:architecture}
\begin{figure}[htbp]
  \centering
  \includegraphics[width=0.75\textwidth]{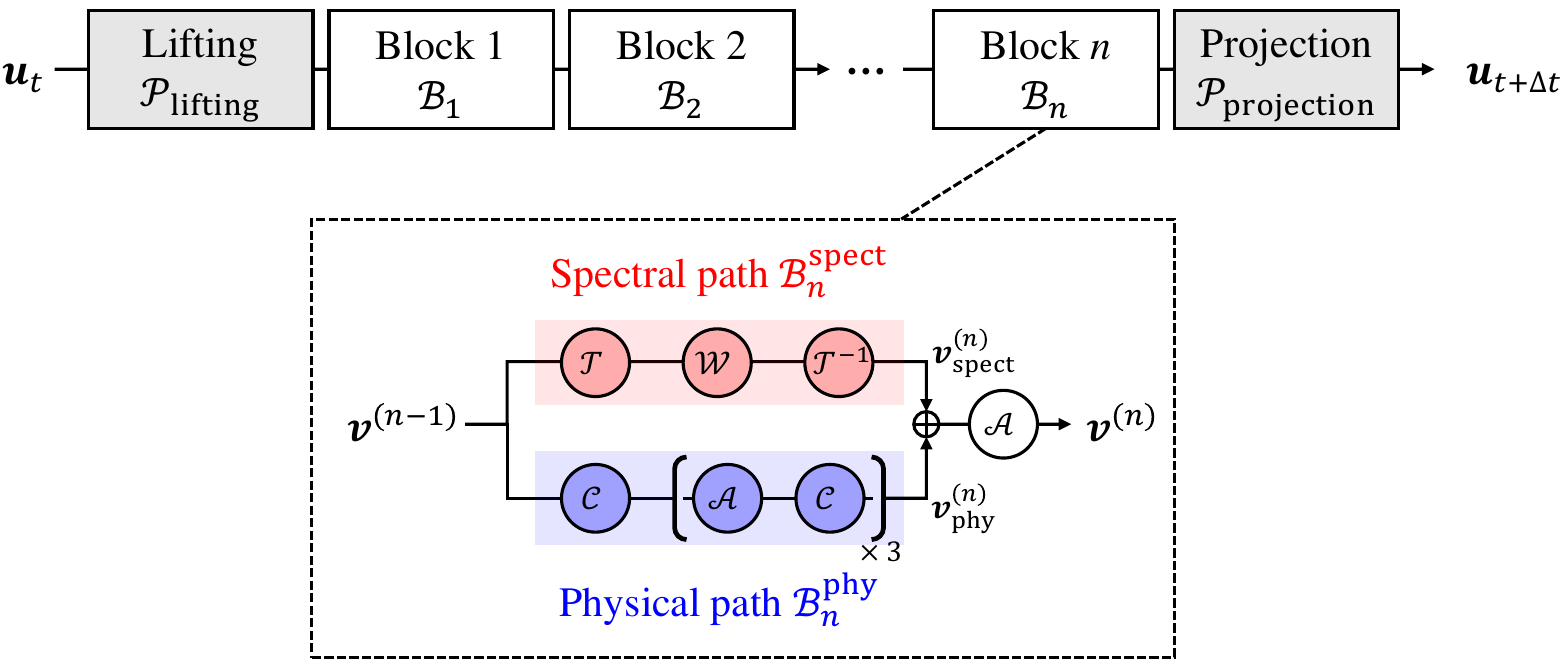}
  \caption{Architecture of local neural operator.}\label{fig_architecture}
\end{figure}

Any architecture of neural networks satisfying Eqs. (\ref{eq:2:lno_definition}-\ref{eq:2:local_related_condition}) can be used for the practice of LNO.
The concrete architecture adopted in this paper is from Ref. \cite{Ye2023b}, 
which comprises a lifting layer $\mathcal{P}_\text{lifting}$, $n$ blocks $\mathcal{B}_i(i=1\sim n)$, and a projection layer $\mathcal{P}_\text{projection}$, as shown in Fig. \ref{fig_architecture}.

When the $4$-channel input $\boldsymbol{u}_t$ is sent to LNO, where the $4$ channels correspond to the $4$ independent variables $\rho$, $T$, $v_x$, and $v_y$, the lifting layer $\mathcal{P}_\text{lifting}$ elevates the number of channels to $40$ to increase the expressibility of the neural network.
The output of the lifting layer is denoted as $\boldsymbol{v}^{(0)}$.
Then, it goes through $n$ blocks with the same architecture but different weights.
A block $\mathcal{B}_i$ includes two parallel paths, the spectral path and the physical path.
The physical path is comprised of $4$ convolutional layers $\mathcal{C}$ with kernel size $3$ and activation function $\mathcal{A}$ (here GELU is adopted) sandwiched between every two convolutional layers.
In the spectral path, the input goes through a $N^{\text{th}}$-order spectral transform $\mathcal{T}$ to be projected to the spectral space, then a linear operation $\mathcal{W}$ followed by the inverse spectral transform $\mathcal{T}^{-1}$.
After the spectral transform, only the lowest $M$ $(M\leq N$) orders of modes are reserved to reduce high-frequency oscillation.
The outputs of the two paths are added and activated as the output of the block.
After these block operations, the final part is the projection layer which turns the number of channels from $40$ to $4$.

Within the architecture, there are several adjustable hyper-parameters, including the number of blocks $n$, the order of spectral transform $N$, the number controlling the stride of the spectral transform $K$, and the order of low-pass filtering in the spectral path $M$.
We set $n=4,N=12,K=2,M=6$, which is the best choice for the current learning problem according to our previous study \cite{Ye2023b}.

\section{Hyper-parameter tuning for backpropagation optimization}
\label{app:hyper_param}
There are 4 main hyper-parameters in the backpropagation optimization, including the optimizer (the learning rate $lr$), the initial value $\Delta \boldsymbol{u}_t^{(0)}$ on the extended domain, the kernel width of the average filter $k$, and the weight decay of the optimizer $\lambda$. 
This appendix investigates the effect of these parameters on the boundary treatment and decides the best choice. 
The flow around a circular cylinder is taken as the object of study with a focus on the boundary imposition of the solid wall. 
The problem settings are consistent with those outlined in Section \ref{sec4:2}.

The principle of hyper-parameter tuning in the boundary treatment is distinct from that in the training of most neural networks on two aspects. 
The first is that, the efficiency occupies a more important position. 
For regular neural networks, the primary goal is to eventually minimize the loss function as much as possible, while the time taken to achieve the minimum is not a major concern. 
However, in boundary optimization, every time step requires a new round of training, making it essential to consider the time cost to prevent significant efficiency loss. 
The second is that, a lower loss does not always indicate a better prediction. 
The loss $\mathcal{L}_\text{BC}$ depends on errors of both the pre-trained LNO and the boundary treatment, schematically formulated as:
\begin{equation}
  \mathcal{L}_\text{BC}=e_\text{LNO}+e_\text{VDE}.
\end{equation}
$e_\text{LNO}$ denotes the error by LNO prediction, assuming the flow field on the extended domain provided by VDE is absolutely accurate; 
the second part $e_\text{VDE}$ refers to the error caused by a nonoptimal flow field from VDE, assuming the LNO prediction is accurate. 
Even VDE successfully finds the accurate flow field $\Delta\boldsymbol{u}$ onn the extended domain ($e_\text{VDE}=0$), $e_\text{LNO}$ can not be removed. 
If $\mathcal{L}_\text{BC}$ is very small (smaller than $e_\text{LNO}$), it suggests VDE has compensated on the searched flow field to make the value of $\mathcal{L}_\text{BC}$ small, i.e., over-fitting occurs. 
In this case, the searched flow field is not the desired accurate one, which reduces $\mathcal{L}_\text{BC}$ but introduces additional errors to the predicted results. 
So, the value of $\mathcal{L}_\text{BC}$ is not the only criteria for the choice of hyper-parameters.

The first two hyper-parameters ($lr$ and $\Delta \boldsymbol{u}_t^{(0)}$) are closely related to the efficiency, so we would like to find their values with highest decline rate of loss function to minimize the number of iterations. 
Meanwhile, the latter two hyper-parameters ($k$ and $\lambda$) connect more with the balance between boundary error and overall performance.
In the discussion, the boundary error and the error contour on the whole computational domain will be considered together to find a moderate choice. 
The default value of hyper-parameters is set as $lr=0.01$, $\Delta \boldsymbol{u}_t^{(0)}=h(\boldsymbol{u}_\text{BC}-\boldsymbol{u}_t)$, $k=3$, $\lambda=1\times{10}^{-4}$. 
When investigating one hyper-parameter, other hyper-parameters are set as the best value if already fine-tuned, otherwise set as the default value.

\subsection{Optimizer (learning rate)}
\begin{figure}[htbp]
  \centering
  \includegraphics[width=0.8\textwidth]{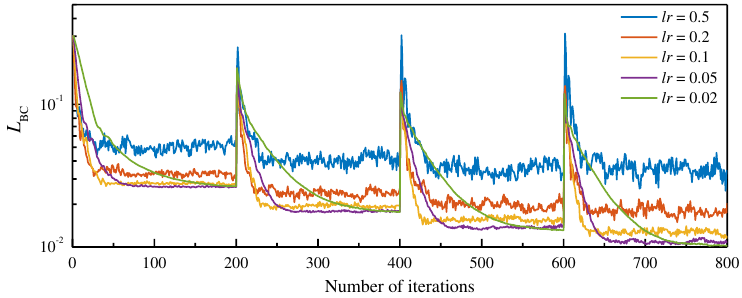}
  \caption{Time history of the loss function $\mathcal{L}_\text{BC}$ with different $lr$. $200$ iterations are executed for one time-marching step, and here totally $4$ steps are shown.}\label{fig_hyper_lr}
\end{figure}

The optimizer Adam \cite{Kingma2015} with different learning rates is discussed here. 
After testing the learning rate in a wide range, Fig. \ref{fig_hyper_lr} exhibits the time history of the loss function $\mathcal{L}_\text{BC}$ across $4$ continuous time steps under five representative learning rates $lr=0.02\sim 0.5$. 
Every time step consists of $200$ rounds of iterations to demonstrate the characteristics of the loss function after both short-term and long-term optimizations. 
In each time step, $\mathcal{L}_\text{BC}$ decreases gradually with the optimization. 
After every $200$ rounds, the optimization progresses to the next time step, where $\mathcal{L}_\text{BC}$ returns to a higher level and then decreases again in the following step of optimization.

When the learning rate is large ($lr=0.2$ and $0.5$), the loss function decreases quickly in the early stages; 
however, this rapid decrease is accompanied by noticeable oscillations, resulting in a higher final value of the loss function. 
Conversely, when the learning rate is small ($lr=0.02$ and $0.05$), the rate of decline (i.e., the slope of the curve) in the early stages is slower, but the curve is smoother, ultimately leading to a lower final value. 
In general, it is best to choose a moderate $lr$ that allows to reach a low value of loss function quickly in a few iterations while minimizing oscillations. 
By this rule, the best choice is $lr=0.1$.

\subsection{Initial value on the extended domain}
Next, we discuss the choice of the initial value $\Delta \boldsymbol{u}_t^{(0)}$ on the extended domain. 
The learning rate $lr$ is set as the found best choice $0.1$. Other two hyper-parameters are set as the default values. 
Option 1 for $\Delta \boldsymbol{u}_t^{(0)}$ is to set
\begin{equation}
  \Delta \boldsymbol{u}_t^{(0)}=h\boldsymbol{u}_t,
\end{equation}
where $h$ is the indicator of whether a point is on the extended domain. Multiplying $h$ emphasizes $\Delta \boldsymbol{u}_t^{(0)}$ only works on the extended domain.
Option 2 is to set
\begin{equation}
  \Delta \boldsymbol{u}_t^{(0)}=h\boldsymbol{u}_\text{BC}.
\end{equation}
The difference between the two options lies in how the interior boundaries, such as the circular cylinder wall, are treated. 
The enclosure of an interior boundary is not subjected to corrosion of domain in LNO process, so there is a distribution of flow field ($\boldsymbol{u}_t$) inside even without boundary imposing. 
Direct setting $\boldsymbol{u}_t$ as the initial value without additional calculation leads to Option 1, whereas assigning the boundary condition value $\boldsymbol{u}_\text{BC}$ on the extended domain leads to Option 2. 
For exterior boundary, there is no $\boldsymbol{u}_t$ on the extended domain, so only Option 1 is not available. 
Option 3 is to overlay a random disturbance $\epsilon\sim N(0,1)$ to Option 2 to imitate the random initialization of weights in neural networks:
\begin{equation}
  \Delta \boldsymbol{u}_t^{(0)}=h(\boldsymbol{u}_\text{BC}+\epsilon).
\end{equation}
Option 4 posits that the flow field in two adjacent time levels changes little, so the ideal $\Delta \boldsymbol{u}_t$ of the two time levels should be similar. 
It takes the final value $\Delta \boldsymbol{u}_{t-\Delta t}^{(s)}$ from the last time step as the initial value of the current step:
\begin{equation}
  \Delta \boldsymbol{u}_t^{(0)}=\Delta \boldsymbol{u}_{t-\Delta t}^{(s)}.
\end{equation}
When calculating the first step, there is no available $\Delta \boldsymbol{u}_{t-\Delta t}^{(s)}$. 
Use Option 2 as the substitute.
The four options are summarized in Table \ref{tab_initial_value_option}.

\begin{table}[htbp]
  \centering
  \caption{Options for initial value on the extended domain}\label{tab_initial_value_option}
  \begin{tabular}{cc}
    \toprule
    Option & Formula\\
    \midrule
    1 & $\Delta \boldsymbol{u}_t^{(0)}=h\boldsymbol{u}_t$\\
    2 & $\Delta \boldsymbol{u}_t^{(0)}=h\boldsymbol{u}_\text{BC}$\\
    3 & $\Delta \boldsymbol{u}_t^{(0)}=h(\boldsymbol{u}_\text{BC}+\epsilon)$\\
    4 & $\Delta \boldsymbol{u}_t^{(0)}=\Delta \boldsymbol{u}_{t-\Delta t}^{(s)}$\\
    \bottomrule
  \end{tabular}
  
\end{table}

\begin{figure}[htbp]
  \centering
  \includegraphics[width=0.8\textwidth]{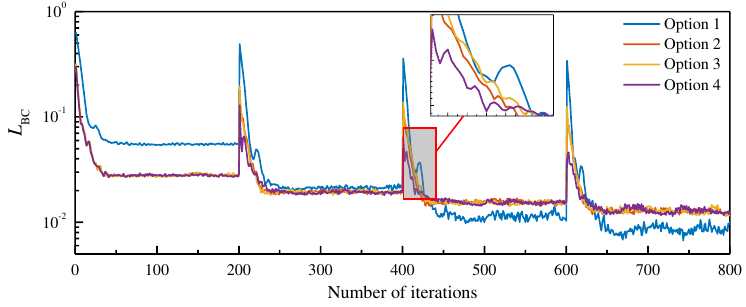}
  \caption{Time history of the loss function $\mathcal{L}_\text{BC}$ with different initial values on the extended domain. $200$ iterations are executed for one time-marching step, and here totally $4$ steps are shown.}\label{fig_hyper_initialvalue}
\end{figure}

Fig. \ref{fig_hyper_initialvalue} shows the time history of the loss function across $4$ continuous time steps under different initial value options. 
In the early stage of optimization, the rate of loss function decline  of each option is similar, but Option 4 can quickly reach a lower level with fewer iterations thanks to a smaller initial loss function value. 
Besides, the loss function of Option 2 and 3 is nearly identical, while Option 1 has a higher value before 25 iterations.
Overall, Option 4 is selected for subsequent calculations due to its superior efficiency and accuracy.

\subsection{Kernel width of the average filter}
\begin{figure}[htbp]
  \centering
  \includegraphics[width=0.8\textwidth]{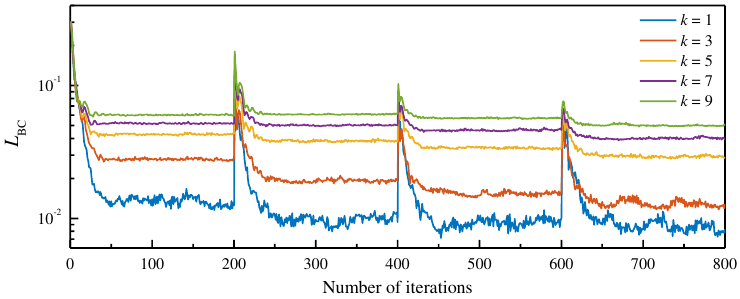}
  \caption{ Time history of the loss function $\mathcal{L}_\text{BC}$ with different kernel width $k$ of the average filter. $200$ iterations are executed for one time-marching step, and here totally $4$ steps are shown.}\label{fig_hyper_k}
\end{figure}

The third hyper-parameter is the kernel width $k$ of the average filter. 
Fig. \ref{fig_hyper_k} illustrates the time history of the loss function across $4$ continuous time steps with $k=1,3,5,7,9$. 
A larger filter width leads to a larger loss function. 
This is because a larger $k$ smooths the flow field on the extended domain and reduces the effectiveness of the optimization process, while a smaller $k$ suggests a more flexible space for optimization. 

\begin{figure}[htbp]
  \centering
  \includegraphics[width=0.9\textwidth]{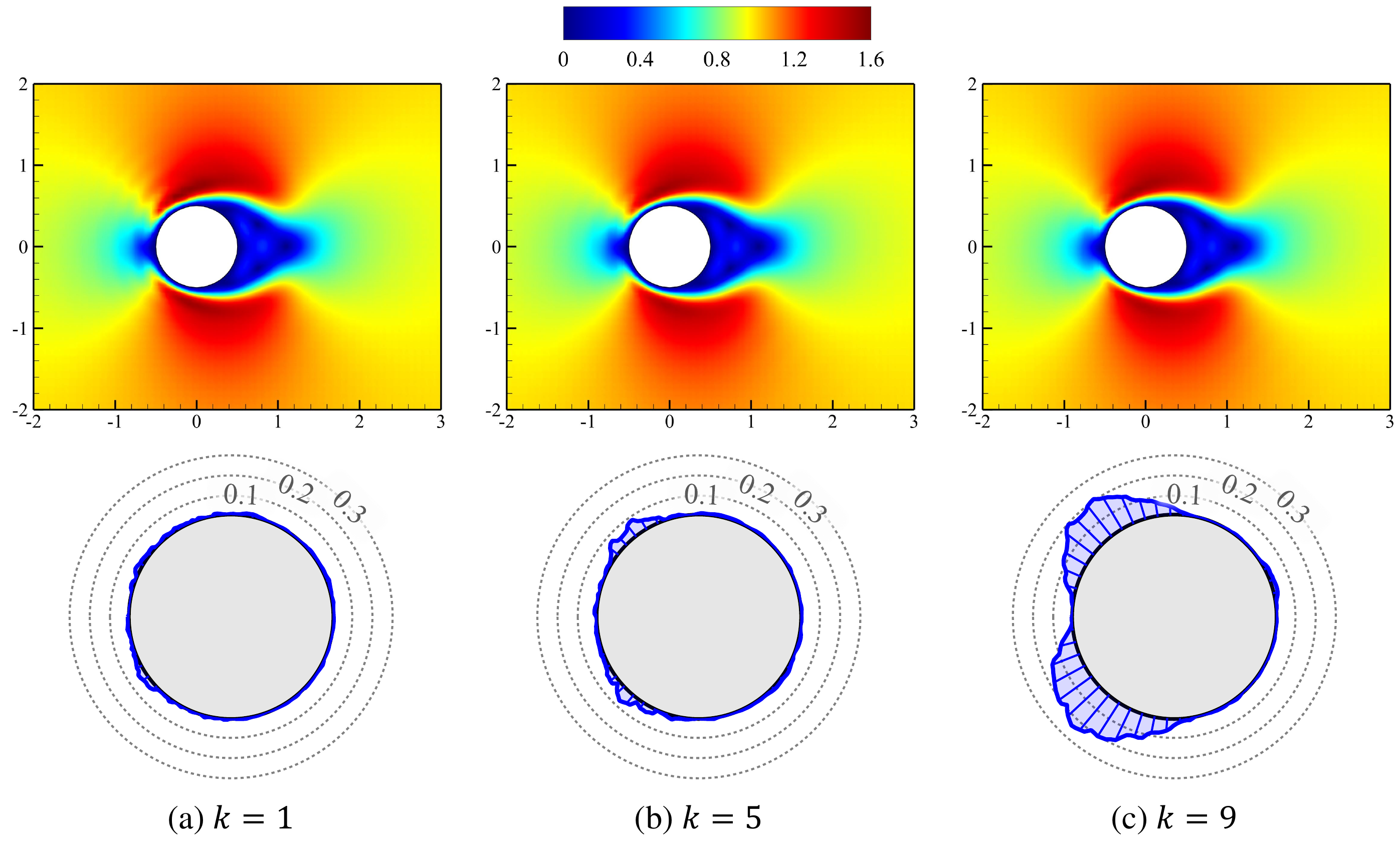}
  \caption{Contours of velocity magnitude and the error distribution on the solid wall for flow around a circular cylinder at $t=1.4$ with different average filters. The three gray dotted circles denote error of magnitude $0.1$, $0.2$, and $0.3$, respectively. Obvious oscillation is observed when $k=1$.}\label{fig_hyper_k_contour}
\end{figure}

However, here the value of loss function is not the primary concern. 
It is found in our experiment that a small $k$ tends to induce oscillations near the boundary, as illustrated in Fig. \ref{fig_hyper_k_contour}. 
This is because each node on the extended domain is relatively independent in the optimization when $k$ is small, the optimizer tries to minimize the boundary loss by all means necessary, which can result in a non-smooth flow field on the extended domain. 
Once the boundary error is below the error of LNO prediction, further reduction of the loss function becomes physically meaningless and causes nonphysical oscillations. 
In the end, $k=5$ is chosen.

\subsection{Coefficient of regularization}
\begin{figure}[htbp]
  \centering
  \includegraphics[width=\textwidth]{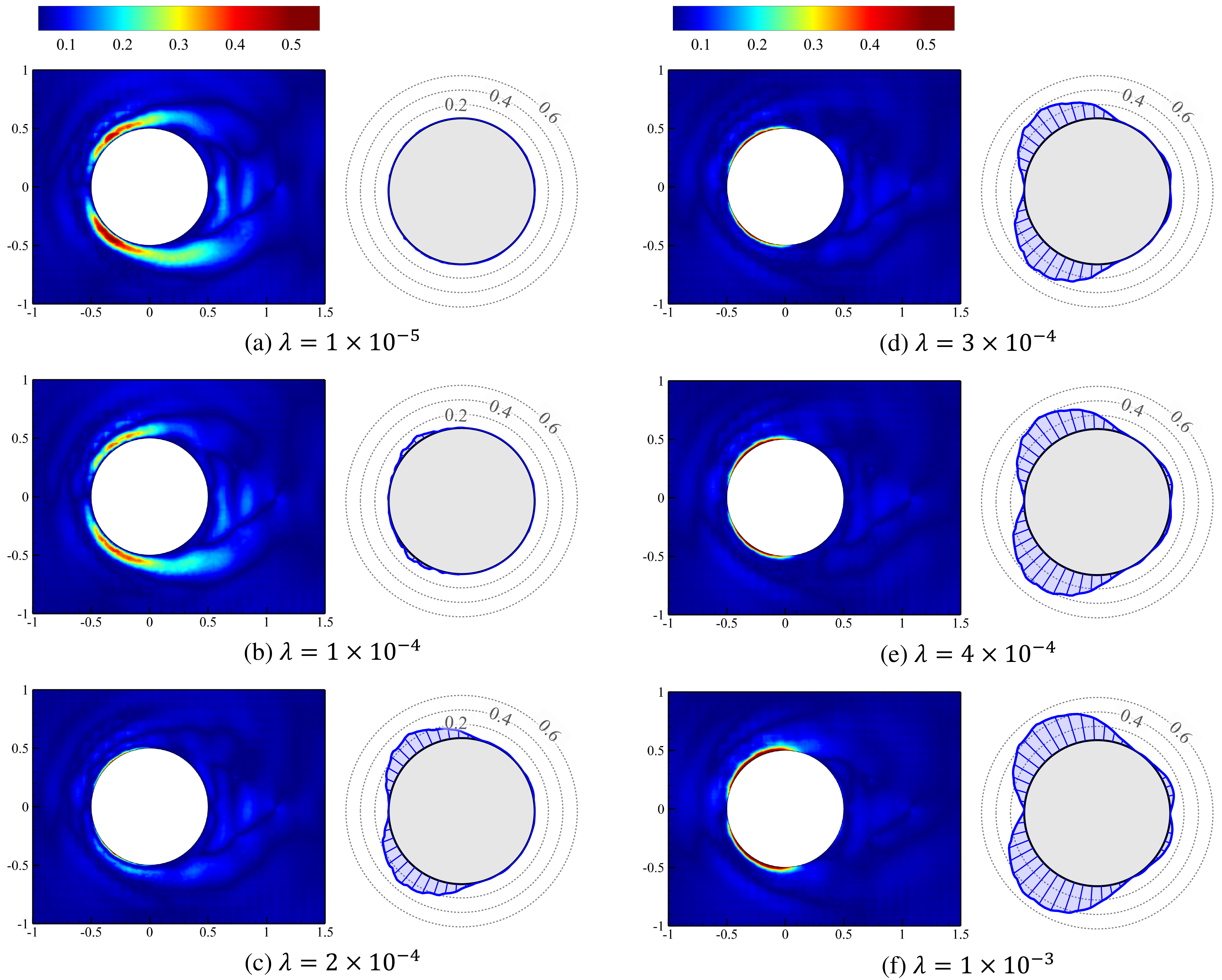}
  \caption{The prediction error with different weight decays. The left column is the contour for error of velocity magnitude, and the right column is the error distribution on the cylinder wall. The three gray dotted circles denote error of magnitude $0.2$, $0.4$, and $0.6$, respectively.}\label{fig_hyper_lambda_contour}
\end{figure}

To prevent over-fitting during the training of neural networks, a regularization term is usually added to the loss function. 
This helps ensure that the optimized weights remain within a reasonable range. 
In the specific context of boundary treatment, it helps to prevent the flow field on the extended domain from being oscillating or overlarge. 
The final loss function for backpropagation is the sum of $\mathcal{L}_\text{BC}$ and the regularization term:
\begin{equation}
  \mathcal{L}=\mathcal{L}_\text{BC}+\lambda \sum_{x\in\Omega_\text{N}}\|\Delta \boldsymbol{u}(\boldsymbol{x})\|^2.
\end{equation}
The coefficient $\lambda$ of the regularization term is named weight decay in PyTorch. 
The common setting for $\lambda$ is ${10}^{-4}\sim{10}^{-3}$. 
A larger $\lambda$ indicates stricter constraint, i.e., the magnitude of $\Delta\boldsymbol{u}$ will be smaller. 
Fig. \ref{fig_hyper_lambda_contour} shows the prediction error of $\lambda$ ranges from $1\times{10}^{-5}$ to $1\times{10}^{-3}$. 
For the case $\lambda=1\times{10}^{-5}$, the boundary condition is well satisfied, but the error on the computational domain is conversely large. 
As $\lambda$ increases, the error on the computational domain reduces and the error concentrates on the boundary. When $\lambda=1\times{10}^{-3}$, the error distribution and the error contour become similar to the result of direct imposition. 
After a tradeoff, a balanced value of $\lambda=2\times{10}^{-4}$ is chosen.

To sum up, the final choice of the hyper-parameters is $lr=0.1$, $\Delta \boldsymbol{u}_t^{(0)}=\Delta \boldsymbol{u}_{t-\Delta t}^{(s)}$, $k=5$, $\lambda=2\times{10}^{-4}$.
The final point to emphasize is that, the elaborate tuning for hyper-parameters in this appendix is a consequence of the inaccucacy of the current pre-trained LNO, rather than a necessity for VDE.
If the accuracy of LNO is improved, the boundary loss $\mathcal{L}_\text{BC}$ could be the only indicator for the tuning, making the process much easier.

\bibliographystyle{elsarticle-num} 
\bibliography{library.bib}

 \end{document}